\PassOptionsToPackage{usenames,dvipsnames}{xcolor}
\documentclass[sigconf, 9pt]{acmart}

\usepackage{balance}

\usepackage{p4testgen_macros}

\acmYear{2023}
\copyrightyear{2023}
\setcopyright{acmlicensed}
\acmConference[ACM SIGCOMM '23]{ACM SIGCOMM 2023 Conference}{September 10--14, 2023}{New York, NY, USA}
\acmBooktitle{ACM SIGCOMM 2023 Conference (ACM SIGCOMM '23), September 10--14, 2023, New York, NY, USA}
\acmPrice{15.00}
\acmDOI{10.1145/3603269.3604834}
\acmISBN{979-8-4007-0236-5/23/09}


\newcommand{\logo}[1]{#1}
\newcommand{\logoheight}{1.1em}

\newcommand{\cornelllogo}{\logo{\includegraphics[height=\logoheight]{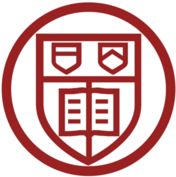}}}
\newcommand{\intellogo}{\logo{\includegraphics[height=\logoheight]{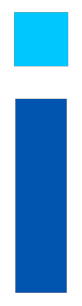}}}
\newcommand{\nyulogo}{\logo{\includegraphics[height=\logoheight]{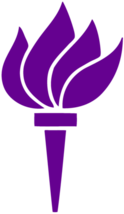}}}
\newcommand{\litsoftlogo}{\logo{\includegraphics[height=\logoheight]{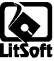}}}
\newcommand{\postmanlogo}{\logo{\includegraphics[height=\logoheight]{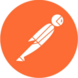}}}
\newcommand{\netdebuglogo}{\logo{$\sharp$}}

\begin{document}
%-------------------------------------------------------------------------------

%\titlenote{Produces the permission block, and copyright information}
%\subtitle{Extended Abstract}

\title{\toolname: An Extensible Test Oracle For \pfour}

\author{
    \texorpdfstring{Fabian Ruffy\textsuperscript{\nyulogo}}{Fabian Ruffy (NYU)}
    \texorpdfstring{Jed Liu\textsuperscript{\postmanlogo}}{Jed Liu (Postman)}
    \texorpdfstring{Prathima Kotikalapudi\textsuperscript{\intellogo}}{Prathima Kotikalapudi (Intel)} 
    \texorpdfstring{Vojt\v{e}ch Havel\textsuperscript{\intellogo}}{Vojt\v{e}ch Havel (Intel} 
    \texorpdfstring{Hanneli Tavante\textsuperscript{\intellogo}}{Hanneli Tavante (Intel)} 
    \texorpdfstring{Rob Sherwood\textsuperscript{\netdebuglogo}}{Rob Sherwood (NetDebug.com)} 
    \texorpdfstring{Vladyslav Dubina\textsuperscript{\litsoftlogo}}{Vladyslav Dubina (Litsoft)} 
    \texorpdfstring{Volodymyr Peschanenko\textsuperscript{\litsoftlogo}}{Volodymyr Peschanenko (Litsoft)}  
    \texorpdfstring{Anirudh Sivaraman\textsuperscript{\nyulogo}}{Anirudh Sivaraman (NYU)} 
    \texorpdfstring{Nate Foster\textsuperscript{\cornelllogo\,\intellogo}}{Nate Foster(Cornell/Intel)}
} 

% \author{
%     \texorpdfstring{Fabian Ruffy\nyulogo}{Fabian Ruffy (NYU)}
%     \texorpdfstring{Jed Liu\postmanlogo}{Jed Liu (Postman)}
%     \texorpdfstring{Prathima Kotikalapudi\intellogo}{Prathima Kotikalapudi (Intel)} 
%     \texorpdfstring{Vojt\v{e}ch Havel\intellogo}{Vojt\v{e}ch Havel (Intel} 
%     \texorpdfstring{Hanneli Tavante\intellogo}{Hanneli Tavante (Intel)} 
%     \texorpdfstring{Rob Sherwood\netdebuglogo}{Rob Sherwood (NetDebug.com)} 
%     \texorpdfstring{Vladyslav Dubina\litsoftlogo}{Vladyslav Dubina (Litsoft)} 
%     \texorpdfstring{Volodymyr Peschanenko\litsoftlogo}{Volodymyr Peschanenko (Litsoft)}  
%     \texorpdfstring{Anirudh Sivaraman\nyulogo}{Anirudh Sivaraman (NYU)} 
%     \texorpdfstring{Nate Foster\cornelllogo~\intellogo}{Nate Foster(Cornell \& Intel)}
% } 

% A hack to insert commata and remove the superscript.
\def \authors{Fabian Ruffy, Jed Liu, Prathima Kotikalapudi, Vojt\v{e}ch Havel, Hanneli Tavante, Rob Sherwood, Vladyslav Dubina, Volodymyr Peschanenko, Anirudh Sivaraman, and Nate Foster}

% The default list of authors is too long for headers}
\renewcommand{\shortauthors}{Ruffy~\etal}

%-------------------------------------------------------------------------------
\begin{abstract}
%-------------------------------------------------------------------------------
% Do not use any macros here so that we can easily paste the abstract.
We present P4Testgen, a test oracle for the P4$_{16}$ language. P4Testgen supports automatic test generation for any P4 target and is designed to be extensible to many P4 targets. It models the complete semantics of the target's packet-processing pipeline including the P4 language, architectures and externs, and target-specific extensions. To handle non-deterministic behaviors and complex externs (e.g., checksums and hash functions), P4Testgen uses taint tracking and concolic execution. It also provides path selection strategies that reduce the number of tests required to achieve full coverage.

We have instantiated P4Testgen for the V1model, eBPF, PNA, and Tofino P4 architectures. Each extension required effort commensurate with the complexity of the target. We validated the tests generated by P4Testgen by running them across the entire P4C test suite as well as the programs supplied with the Tofino P4 Studio. Using the tool, we have also confirmed 25 bugs in mature, production toolchains for BMv2 and Tofino.
\end{abstract}

\maketitle

{\let\thefootnote\relax
\makeatletter\def\Hy@Warning#1{}\makeatother
\footnotetext{
    \textsuperscript{\nyulogo}New York University
    \textsuperscript{\postmanlogo}Postman
    \textsuperscript{\intellogo}Intel
    \textsuperscript{\netdebuglogo}NetDebug.com
    \textsuperscript{\litsoftlogo}Litsoft
    \textsuperscript{\cornelllogo}Cornell University
}
}

\section{Introduction}
\label{sec:introduction}
We present \toolname, an extensible \emph{test oracle} for the \pfoursixteen~\cite{p416_spec} language. Given a \pfour program and sufficient time, it generates an exhaustive set of tests that cover every reachable statement in the program. Each test consists of an input packet, control-plane configuration, and the expected output packet. 

\toolname generates tests to validate the \emph{implementation} of a \pfour program. Such tests ensure that the device executing the \pfour code (commonly referred to as ``target'') and its toolchain (\ie the compiler~\cite{p416}, control-plane~\cite{p4runtime_spec, tdi}, and various API layers~\cite{sai, sonic, pins}) implement the behaviors specified by the \pfour program.

Tests generated by \toolname can be used by manufacturers of \pfour-programmable equipment to validate the toolchains associated with their equipment~\cite{tofino, ipu, nvidia_doca_sdk,pensando_p4, keysight, silicon_one, marvell}, by \pfour compiler writers for debugging optimizations and code transformations~\cite{p416, oxide_compiler}, and by network owners to check that both fixed-function and programmable targets implement behaviors as specified in \pfour, including standard and custom protocols~\cite{meissa, switchv}.

The idea of generating an exhaustive set of tests for a given \pfour program is not new. But prior work has largely focused on a specific \pfour architecture~\cite[\S4]{p416_spec}. For example, \texttt{p4pktgen}~\cite{p4pktgen} targets \bmv~\cite{bmv2}, Meissa~\cite{meissa} and \texttt{p4v}~\cite{p4v} target Tofino~\cite{tofino}, and SwitchV~\cite{switchv} targets fixed-function switches. The primary reason why these tools are so specialized is development effort. Building \pfour validation tools requires simultaneously understanding (i) the \pfour language, (ii) formal methods, and (iii) target-specific behaviors and quirks. Finding developers that satisfy this trifecta even for a single target is already challenging. Finding developers that can design a general tool for all targets is even harder. The unfortunate result is that developer effort has been fragmented across the \pfour ecosystem. Most \pfour targets today lack adequate test tooling, and advances made with one tool are difficult to port over to other tools.

Our position is that this fragmentation is undesirable and entirely avoidable. While there may be scenarios that warrant the development of target-specific tools, in the common case---\ie generating \inout pairs for a given program---the desired tests can be derived from the semantics of the \pfour language, in a manner that is largely decoupled from the details of the target. Developing a common, open-source platform for validation tools has several benefits. First, common software infrastructure (lexer, parser, type checker, etc.) and an interpreter that realizes the core \pfour language semantics can be implemented just once and shared across many tools. Second, because it is open-source, improvements can be contributed back to \toolname and benefit the whole community. 

\toolname combines several techniques in an open-source tool suitable for production use. First, \toolname provides an extensible framework for defining the  semantics of the whole program (``whole-program semantics''), combining the semantics of the \pfour code along with the semantics of the target on which it is executed. A \pfour program generally consists of several \pfour blocks (with semantics provided by the language specification) that are separated by interstitial architecture-specific elements (with semantics provided by the target). \toolname is the first tool that provides an extensible framework for such whole-program semantics, using a carefully designed interpreter based on the open-source \pfour compiler (\pfourc)~\cite{p416}. Second, while \toolname ultimately uses an SMT solver to generate tests, it also handles the ``awkward squad'' of complex functions that are difficult to model using an SMT solver---\eg checksums, undefined values, randomness, and so on. To achieve this, \toolname uses taint tracking, concolic execution, and a precise model of packet sizing to model the semantics of the program accurately and at bit-level granularity. Third, \toolname offers advanced path selection strategies that can efficiently generate tests that achieve full statement coverage, even for large \pfour programs that suffer from path explosion. In contrast to prior work, these strategies are fully automated and do not require annotations to use effectively.

To recap, \toolname's key technical innovations are as follows:
\begin{CompactEnumerate}
\item \textbf{Whole-program semantics}: Most \pfour targets perform processing that is not defined by the \pfour program itself and is target-specific. \toolname uses \emph{pipeline templates} to succinctly describe the behavior of an entire pipeline as a composition of \pfour-programmable blocks and interstitial target-specific elements.
\item \textbf{Target-specific extensions}: Many real-world \pfour targets deviate from the \pfoursixteen specification in ways small and large. To accommodate these deviations, \toolname's extensible interpreter supports target-specific \emph{extensions} to override default \pfour behavior, including initialization semantics and an intricate model of \emph{packet-sizing}, which accommodates targets that modify packet sizes during processing.
\item \textbf{Taint analysis}: Targets can exhibit non-deterministic behavior, making it impossible to predict test outcomes. To ensure that generated tests are reliable, \toolname uses \emph{taint analysis} to track non-deterministic portions of test outputs.
\item \textbf{Concolic execution}: Some targets have features that can not easily be modelled using an SMT solver. \toolname uses \emph{concolic execution}~\cite{dart,cute} to model features such as hash functions and checksums.
\item \textbf{Path selection strategies:} Real-world \pfour programs often have a huge number of paths, making full path coverage infeasible. \toolname provides heuristic \emph{path selection strategies} that can achieve full statement coverage, usually with orders of magnitude fewer tests than other approaches.
\end{CompactEnumerate}

% This is OK: We validated our design in this particular way, and we found bugs!
To validate our design for \toolname, we instantiated it for \numextensions different real-world targets and their corresponding \pfour architecture: the \vonemodel~\cite{bmv2_targets} architecture for \bmv, the \ebpfmodel~\cite{p4c-xdp} architecture for the Linux kernel~\cite{ebpf}, the \pna~\cite{pna_spec} architecture for the DPDK SoftNIC~\cite{dpdk_softnic}, the \tna~\cite{open_tofino} architecture for the Tofino 1 chip~\cite{tofino}, and the \ttwona architecture for the Tofino 2 chip~\cite{tofino2}. All \numextensions instantiations implement whole-program semantics without requiring modification to the core parts of \toolname. We have tested the correctness of the \toolname oracle itself by generating \inout tests for example \pfour programs of all listed architectures. Executing \toolname's tests using the appropriate target toolchains, we have found \tnabugsconfirmed bugs in the toolchain of the Tofino compiler and \bmvbugsconfirmed in the toolchain of \bmv. \toolname is available at the following URL: \url{https://p4.org/projects/p4testgen}.

\newcounter{challenges}

%-------------------------------------------------------------------------------
\section{Motivation and Challenges}
%-------------------------------------------------------------------------------
\label{sec:motivation}
\pfour offers new capabilities for specifying network behavior, but this flexibility comes at a cost: network owners must now navigate toolchains that are larger and more complex than with fixed-function devices. So, as the \pfour ecosystem matures, increased focus is being placed on tools for validating \pfour implementations~\cite{meissa,switchv,p4pktgen,pta,fp4,dbval,gauntlet,petr4}, often by exercising \inout tests.

At first glance, the task of generating tests for a given \pfour program may seem relatively straightforward. Prior work such as \texttt{p4pktgen}~\cite{p4pktgen}, \texttt{p4v}~\cite{p4v}, P4wn~\cite{p4wn}, Meissa~\cite{meissa}, and SwitchV~\cite{switchv} has shown that it is possible to automatically generate tests using techniques from the programming languages and software engineering literature~\cite{king76,cute,dart}. The precise details vary from tool to tool, but the basic idea is to first use symbolic execution to traverse a path in the program, collecting up a symbolic environment and a path constraint, and then use a first-order theorem prover (i.e., SAT/SMT solver) to compute an executable test. The theorem prover fills in the \inout packet(s) from the symbolic environment to satisfy the path constraint and also computes control-plane configurations required to execute the selected path---\eg forwarding entries in match-action tables.

\Para{Technical Challenges.} While prior work has shown the feasibility of automatic test generation using symbolic execution, existing tools have focused on specific targets (\eg Tofino) and abstracted away important details (\eg non-standard packets and other ``corner cases'' in the language), which limits their applicability in practice. In contrast, our goal for \toolname is to develop a general and extensible test oracle for \pfour that can be readily applied to real-world \pfour programs on arbitrary targets. Achieving this goal requires overcoming several technical challenges, described below.

\Para{(1) Missing inter-block semantics.}
\refstepcounter{challenges}\label{para:challenge_1}
A \pfour program only specifies the target behavior \emph{within} the \pfour programmable blocks in the architecture. It does not specify the execution order of those blocks, or how the output of one block feeds into the input of the next, \ie target-specific semantics in the interstices between blocks. For instance, Tofino's \tna and \ttwona architectures contain independent ingress and egress pipelines, with a traffic manager between them. The traffic manager can forward, drop, multicast, clone, or recirculate packets, depending on their size, content, and associated metadata. As another example, the \pfour specification states that, if extracting a header fails because the packet is too short, the parser should step into \texttt{reject} and exit~\cite[\S12.8.1]{p416_spec}. However, the semantics after exiting the \texttt{reject} state is left up to the target: some drop the packet, others consider the header uninitialized, while others silently add padding to initialize the header. None of these behaviors are captured by the \pfour program itself. \toolname offers features for describing such \emph{inter-block semantics} (\S \ref{sec:design_whole_program}).

\Para{(2) Target-specific intra-block semantics.}\refstepcounter{challenges}\label{para:challenge_2}
Even though \pfour describes the behavior of a programmable block, targets may also have different \emph{intra-block semantics}, \ie they interpret the \pfour code within the programmable block differently. The \pfour specification delegates numerous decisions to targets and targets may not implement all parts of the specification. For instance, hardware restrictions can make it difficult to implement parser exceptions faithfully~\cite{tofino_spec_deviation}. Match-action table execution can also be customized using target-specific properties (\eg action profiles) and annotations can influence the semantics of headers and other language constructs in subtle ways---see Tbl.~\ref{tbl:target_quirks} in the appendix for a (non-exhaustive) list of target-specific deviations. As part of its whole-program semantics model, \toolname offers a flexible abstract machine based on an extensible class hierarchy, which makes it easy to accommodate target-specific refinements of the \pfour specification.

\Para{(3) Unpredictable program behavior.}
\refstepcounter{challenges}\label{para:challenge_3}
Not all parts of a \pfour program are well-specified by the code. For instance, reading from an uninitialized variable may return an undefined value. \pfour programs may also invoke arbitrary extern functions, such as pseudo-random number generators, which produce unpredictable output. To ensure that generated tests are deterministic, \toolname needs facilities to track program segments that may cause unpredictable output. \toolname uses \emph{taint-tracking} to keep track of unpredictable bits in the output (\S \ref{sec:design_taint}), ensuring that it never produces nondeterministic tests unless explicitly asked to do so.

\Para{(4) Complex primitives.}
\refstepcounter{challenges}\label{para:challenge_4}
Like other automated test generation tools, \toolname relies on a first-order theorem prover to compute \inout tests. However, not all primitives can easily be encoded into first-order logic---\eg checksums and other hash functions, or programs that modify the size of the packet using dynamic values. For instance, consider a program that uses the \texttt{advance} function to increment the parser cursor by an amount that depends on values within the symbolic input header. Modeling this behavior precisely either requires bit vectors of symbolic width, which is not well-supported in theorem provers, or branching on every possible value, which is impractical. \toolname uses \emph{concolic execution} to accommodate computations which cannot be encoded into first-order logic (\S \ref{sec:design_concolic}).

\Para{(5) Path explosion.}
\refstepcounter{challenges}\label{para:challenge_5}
By default, \toolname uses depth-first search (\dfscov) to select paths throughout the \pfour program. It does not prioritize any path and it explores all valid paths to exhaustion.  However, real-world \pfour programs often have dense parse graphs and large match-action tables, so the number of possible paths grows exponentially~\cite{p4v,vera}. Achieving full path coverage would require generating an excessive number of tests. \toolname provides \emph{strategies} for controlling the selection of paths, including random strategies and coverage-guided heuristics that seek to follow paths containing previously unexplored statements. These strategies enable achieving full statement coverage with orders of magnitude fewer tests compared to other approaches (\S \ref{sec:useful_tests}).

\Para{Outlook.}
To our knowledge, \toolname is the first test generation tool for \pfour that meets all of these challenges. Moreover, \toolname has been designed to be fully extensible, and it is freely available online under an open-source license, as a part of the \pfourc compiler framework. We are hopeful that \toolname will become a valuable resource for the \pfour community, providing the  necessary infrastructure to rapidly develop accurate test oracles for a wide range of \pfour architectures and targets, and generally reducing the cost of designing, implementing, and validating data planes with \pfour.
%---- FIGURE ----%
\begin{figure}[t]
    \centering
    \includegraphics[width=\columnwidth]{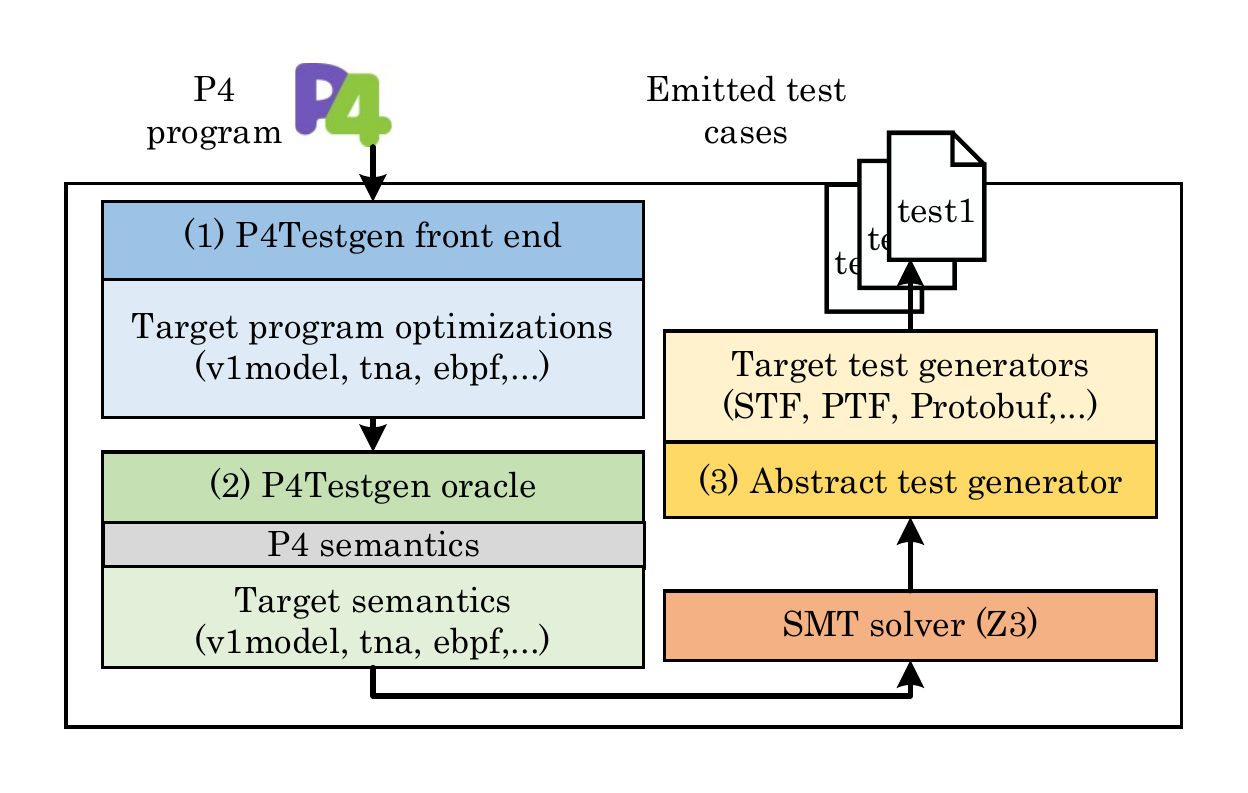}
    % \vspace{-2.5em}
    \caption{The \toolname test case generation process.}
    \label{fig:testgen_workflow}
\end{figure}
%---- FIGURE ----%

\section{\toolname Overview}
\label{sec:design_overview}
As shown in Fig.~\ref{fig:testgen_workflow}, \toolname generates tests using symbolic execution. It selects a path in the program, encodes the associated path constraint as a first-order formula, and then solves the constraint using an SMT solver. If it finds a solution to the constraint, then it emits a test comprising an input packet, output packet(s), and any control-plane configuration required to execute the path. If it finds no solution, then the path is infeasible. Along with the generated tests, \toolname reports which segments of the program (statements, externs, actions) are covered by each test. \toolname's workflow can be summarized as a three-step process.

\Para{Step 1: Translate the input program and target into a symbolically executable representation.}
\toolname takes as input a \pfour program, the target architecture, and the desired test framework (\eg STF~\cite{p4_slides} or PTF~\cite{ptf}). It parses the \pfour program and converts it into the \pfourc intermediate representation language (P4C-IR). \toolname then transform the parsed P4C-IR into a simplified form that makes symbolic execution easier, \eg \toolname unrolls parser loops and replaces run-time indices for header stacks with conditionals and constant indices. The correctness of \toolname's tests is predicated on the correctness of the \pfourc front-end and these transformations.

\Para{Step 2. Generate the test case specification.}
After the input program has been parsed and transformed, \toolname symbolically executes the program by stepping through individual AST nodes (parser states, tables, statements). By default, the \toolname interpreter provides a reference implementation for each \pfour construct. However, each step can be customized to reflect target-specific semantics by overriding methods in the symbolic executor. Targets must also define whole-program semantics (\S\ref{sec:design_whole_program}) which describe how individual \pfour blocks are chained together (\ie the order in which a packet traverses the \pfour blocks), what kind of parsable data can be appended or prepended to packets (\eg frame check sequences), and how target system data (also called intrinsic metadata) is initialized. Typically, this target-specific information can be inferred from the documentation for the \pfour architecture or the target itself. Detailed knowledge of hardware microarchitecture is not necessary. 

\Para{Step 3. Emit the test case.}
Once \toolname has executed a path, it emits an abstract test specification, which describes the expected system state (\eg registers and counters) and output packet(s) for the given packet input and control-plane configuration. This abstract test specification is then concretized for execution on different test frameworks (STF, PTF, etc.).

%---- FIGURE ----%
\renewcommand{\theFancyVerbLine}{\textcolor[rgb]{0.497495,0.497587,0.497464}{\fontsize{8}{8}\arabic{FancyVerbLine}}}
\begin{figure}[ht]
% % % % % % % % % % % % % % % % % % % % % % % % % % % % % % % % % 
% FIGURE 1
\captionsetup[subfigure]{aboveskip=3pt,belowskip=3pt}
\centering

\begin{subfigure}[]{\columnwidth}
\begin{minted}[frame=bottomline, numbers=left, fontsize=\fontsize{8}{8}\ttfamily, numbersep=3pt]{text}
parser Parser(...) {
    pkt.extract(hdr.eth);
    transition accept;
}
control Ingress(...) {
    action set_out(bit<9> port) {
        meta.output_port = port;
    }
    table forward_table {
        key = { h.eth.type: exact; @name("type") }
        actions = { noop; // Default action.
                    set_out; }
    }
    h.eth.type = 0xBEEF;
    forward_table.apply();
}
\end{minted}
\begin{minted}[frame=bottomline, numbers=none, fontsize=\fontsize{8}{8}\ttfamily, numbersep=3pt]{text}
        Size Port eth.dst      eth.src      eth.type
--- Test 1 -------------------------------------------------
Input:  112  0    000000000000 000000000000 0000    
Output: 112  0    000000000000 000000000000 BEEF    
--- Test 2 -------------------------------------------------
Input:  112  0    000000000000 000000000000 0000    
Output: 112  2    000000000000 000000000000 BEEF     
Table Config: match(type=0xBEEF),action(set_out(2))
--- Test 3 -------------------------------------------------
Input:  112  0    000000000000 000000000000 0000                   
Output: 112  0    000000000000 000000000000 BEEF                   
Table Config: match(type=0xBEEF),action(noop())    
--- Test 4 -------------------------------------------------
Input:  96   0    000000000000 000000000000          
Output: 96   0    000000000000 000000000000          
\end{minted}
\caption{\pfour program that forwards using the source MAC.}
\label{fig:interpreter_example_a}
\end{subfigure}

\begin{subfigure}[]{\columnwidth}
\begin{minted}[frame=bottomline, numbers=left, fontsize=\fontsize{8}{8}\ttfamily, numbersep=3pt]{text}
parser Parser(...) {
    pkt.extract(hdr.eth);
    transition accept;
}
control Verify(...) {
    meta.checksum_err = verify_checksum(
    hdr.eth.isValid(),
    {hdr.eth.dst, hdr.eth.src}, 
    hdr.eth.type); 
}
control Ingress(...) {
    if (meta.checksum_err == 1) {
        mark_to_drop(); // Drop packet.
    }
}
\end{minted}
\begin{minted}[frame=bottomline, numbers=none, fontsize=\fontsize{8}{8}\ttfamily, numbersep=3pt]{text}
        Size Port eth.dst      eth.src      eth.type
--- Test 1 -------------------------------------------------
Input:  112  0    BADC0FFEE0DD F00DDEADBEEF
Output: 112  0    BADC0FFEE0DD F00DDEADBEEF
--- Test 2 -------------------------------------------------
Input:  112  0    BADC0FFEE0DD F00DDEADBEEF FFFF
--- Test 3 -------------------------------------------------
Input:  112  0    BADC0FFEE0DD F00DDEADBEEF 7072
Output: 112  0    BADC0FFEE0DD F00DDEADBEEF 7072
\end{minted}
\caption{\pfour program that validates the Ethernet checksum.}
\label{fig:interpreter_example_b}
\end{subfigure}
\caption{\toolname test case examples. ``Port'' denotes the \inout port. ``Size'' is the packet bit-width.}

\label{fig:interpreter_examples}
% \vspace{-0.1em}
\end{figure}
%---- FIGURE ----%

%-------------------------------------------------------------------------------
\subsection{\toolname in Action}
%-------------------------------------------------------------------------------
\label{sec:examples}
As an example to illustrate the use of \toolname, consider two \pfour programs, as shown in Fig.~\ref{fig:interpreter_examples}, written for a fictitious, \bmv-like target with a single parser and control block.

\Para{Example 1.}
In the first program (Fig.~\ref{fig:interpreter_example_a}), Ethernet packets are forwarded based on a table that matches on the EtherType field. There are four different \inout pairs that could be generated. The first pair is a valid Ethernet packet, but no table entries are associated with the input. Since the default action is \texttt{noop}, the output port of the packet does not change. The second pair is a configuration with a table entry that executes \texttt{set\_out} whenever \texttt{h.eth.type} matches a given value. Since the program previously set \texttt{h.eth.type} to \texttt{0xBEEF} the table entry must match on \texttt{0xBEEF}. The output port is defined by the control plane. The third pair is similar, except \texttt{noop} is chosen as action, which does not alter the output port. For the last input pair the packet is too short and the \texttt{extract} call fails. Hence, the target stops parsing and continues to the control. For this particular target the packet will be emitted, but \texttt{forward\_table} will not execute because the match key is uninitialized. \toolname is able to generate four distinct tests for this program. For \inout pairs 2 and 3, \toolname synthesizes control plane entries, which execute the appropriate action. For \inout pair 4, \toolname makes use of its \emph{packet sizing} (\S~\ref{sec:design_packet_sizing}) implementation to generate a packet that is too short. \toolname uses \emph{taint tracking} (\S~\ref{sec:design_taint}) to identify that \texttt{h.eth.type} is uninitialized. Since this target will not match on uninitialized keys, \toolname does not generate an entry for \texttt{forward\_table}.

\Para{Example 2.}
The second program (Fig.~\ref{fig:interpreter_example_b}) parses an Ethernet header. If it is valid (line 7), the program tests whether the checksum computed on \texttt{hdr.eth.dst} and \texttt{hdr.eth.src} (lines 6--9) corresponds to the value in field \texttt{hdr.eth.type} (line 10).\footnote{Note this is a non-standard use of EtherType for the sake of the example.} If not, \texttt{meta.checksum\_err} is set to \texttt{true} and the packet is dropped. This program produces three distinct \inout pairs. The first pair is an input packet that is too short, which causes the Ethernet header to be invalid. Hence, \texttt{verify\_checksum} is not executed, the error is not set, and the packet is forwarded. The second and third \inout pair include a valid Ethernet header. In the second pair, \texttt{hdr.eth.type} matches the computed checksum value and the packet is forwarded. In the third pair, the value does not match and the packet is dropped. Note that for \inout pair 2 and 3, \toolname uses \emph{concolic execution} (\S~\ref{sec:design_concolic}) to model the checksum computation. \toolname picks a random concrete assignment to \texttt{hdr.eth.dst} and \texttt{hdr.eth.src}, computes the checksum, and compares the result to \texttt{hdr.eth.type}. As there are no other restrictions on the value of \texttt{hdr.eth.dst} and \texttt{hdr.eth.src}, \toolname produces tests where the checksum either matches (test 3) or does not match (test 2).

\Para{Summary.} As shown, \toolname prefers to maximize program coverage even though it may lead to path explosion. The behaviors exhibited by the tests in Fig.~\ref{fig:interpreter_examples} are possible on the underlying targets and testing them is important. Indeed, we have used \toolname to uncover a variety of bugs in compilers, drivers, and software models---see \S\ref{sec:evaluation} for details. Moreover, these bugs were not for toy programs or early versions of systems under development. Rather, they were found in production code for mature systems that had already undergone extensive validation with traditional testing.

%-------------------------------------------------------------------------------
\section{Whole-Program Semantics}
%-------------------------------------------------------------------------------
\label{sec:design_whole_program}
The symbolic execution of \pfour programs requires a model of not only the \pfour code blocks (parsers, controls, etc.), but also the transformations performed by the rest of the target. However, the \pfour language does not specify the behavior of the target architecture (e.g., the order of execution of \pfour programmable blocks). \toolname addresses this limitation through a \emph{flexible abstract machine} and \emph{pipeline templates}. 

\subsection{\toolname's Abstract Machine}
\label{sec:design_cps}
Fig.~\ref{fig:abstract_machine} summarizes the design of the abstract machine that powers \toolname's symbolic executor. It has standard elements, such as a stack frame, symbolic environment, and so on, as well as a continuation, which encodes the rest of the computation. A full treatment of continuations~\cite{cps} is beyond the scope of this paper. In a nutshell, continuations make it easy to encode non-linear control flow such as packet recirculation, which many P4 architectures support, and they also preserve execution contexts across paths, which is helpful for implementing different path selection heuristics. 

%---- FIGURE ----%
\begin{figure}[t]
\begin{minted}[frame=bottomline, fontsize=\fontsize{8}{8}\ttfamily]{c++}
class ExecutionState {
   // Small-step Evaluator: can be overriden by targets
   friend class SmallStepEvaluator; 
   // Symbolic Environment: maps values to variables
   SymbolicEnv env;
   // Visited: previously-visited nodes for coverage
    P4::Coverage::CoverageSet visitedNodes;
   // Path Constraint: must be satified to execute this path
   std::vector<const IR::Expression *> pathConstraint;
   // Stack: tracks namespaces, declarations, and scope
   std::stack<const StackFrame &>> stack;
   // Continuation: remainder of the computation
   Continuation::Body body;
   ...
}
\end{minted}
\caption{Execution state for \toolname's abstract machine.}
\label{fig:abstract_machine}
\end{figure}
%---- FIGURE ----%

\subsection{The Pipeline Template}
Pipeline templates are a succinct mechanism for describing the \emph{pipeline state} and \emph{control flow} for an architecture---and with those two, its inter-block semantics. By default, they capture the common case where the state associated with the packet simply flows between \pfour-programmable blocks in a straightforward manner---\eg by copying output variables of one block to the input variables of the next. \toolname also handles more complicated forms of packet flow in the architecture, such as recirculation, but this requires writing explicit code against the abstract machine.

\subsubsection{Pipeline State}
Pipeline state describes the per-packet data that is transferred between \pfour-programmable blocks. Fig.~\ref{fig:target_state} shows the pipeline state description for the \vonemodel in a simple C++ DSL. The objects listed in the data structure are mapped onto the programmable blocks in the top-level declaration of a \pfour program (shown in comments). The declaration order of these objects determines the order in which the blocks are executed by default, but this can be overridden by the pipeline control flow based on a packet's per-packet data values. Arguments with the same name are threaded through the programmable blocks in execution order. For example, the \texttt{*hdr} parameter in the parser is first set undefined, as it is used in an \texttt{out} position as seen by the comments in Fig.~\ref{fig:target_state}. After executing the parser, it is copied into the checksum unit, then to the ingress control, etc.

%---- FIGURE ----%
\begin{figure}[t]
\begin{minted}[frame=bottomline, fontsize=\fontsize{8}{9}\ttfamily]{c++}
ArchitectureSpec("V1Switch", {
  // parser Parser<H, M>(packet_in b,
  //                     out H parsedHdr,
  //                     inout M meta,
  //                     inout standard_metadata_t sm);
  {"Parser", {none, "*hdr", "*meta", "*sm"}},
  // control VerifyChecksum<H, M>(inout H hdr,
  //                              inout M meta);
  {"VerifyChecksum", {"*hdr", "*meta"}},
  // control Ingress<H, M>(inout H hdr,
  //                       inout M meta,
  //                       inout standard_metadata_t sm);
  {"Ingress", {"*hdr", "*meta", "*sm"}},
  // control Egress<H, M>(inout H hdr,
  //            inout M meta,
  //            inout standard_metadata_t sm);
  {"Egress", {"*hdr", "*meta", "*sm"}},
  // control ComputeChecksum<H, M>(inout H hdr,
  //                       inout M meta);
  {"ComputeChecksum", {"*hdr", "*meta"}},
  // control Deparser<H>(packet_out b, in H hdr);
  {"Deparser", {none, "*hdr"}}});
\end{minted}
\caption{The pipeline state for the \texttt{v1model} architecture. Comments describe the associated \pfour block. The word \texttt{none} indicates parameters irrelevant to the state.}
\label{fig:target_state}
\end{figure}
%---- FIGURE ----%

%---- FIGURE ----%
\begin{figure}[t!]
% FIGURE 1
\begin{subfigure}[h]{\columnwidth}
\begin{lstlisting}[style=P4Style]
control Ingress(...) {
    if (hdr.ip.ttl == 0) {
        m.drop = 1; // Drop packet
    }
    if (hdr.ip.ttl == 1) {
        resubmit.emit(m); // Resubmit packet
    }
}
Pipeline(I_Parser(), Ingress(), I_Deparser(), E_Parser(), Egress(), E_Deparser()) pipe;
\end{lstlisting}
% \vspace{-1.5em}
\caption{\pfour program snippet that sets metadata state.}
\label{fig:control_flow_a}
\end{subfigure}
% FIGURE 2
\begin{subfigure}[b]{\columnwidth}
\centering
\includegraphics[width=\linewidth]{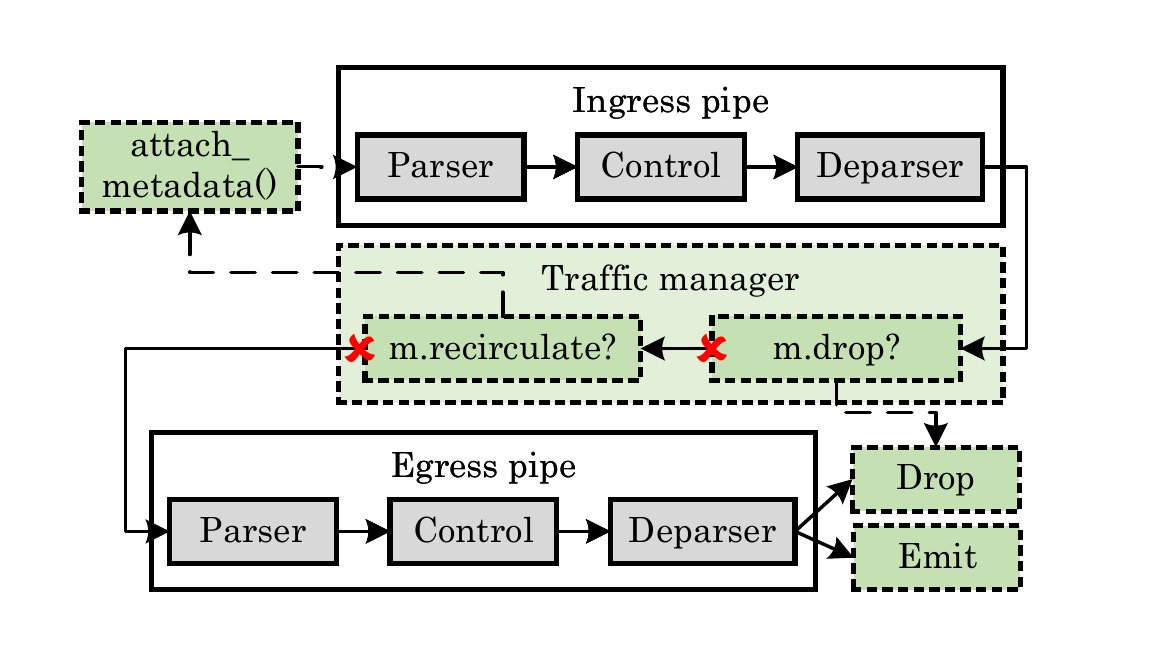}
% \vspace{-2.5em}
\caption{\toolname control-flow. Dashed segments are target-defined. \textcolor{red}{\ding{55}} is \texttt{false}}
\label{fig:control_flow_b}
\end{subfigure}
\caption{\toolname's pipeline control flow.}
\label{fig:control_flow}

\end{figure}

%---- FIGURE ----%

\subsubsection{Pipeline Control Flow}

\toolname allows extension developers to provide code to model arbitrary interpretation of the pipeline state. Fig.~\ref{fig:control_flow} shows an example of a \pfour program snippet being interpreted in the context of \toolname's pipeline control flow. The target is a fictitious target with an implicit traffic manager between ingress and egress pipelines. The green dashed segments in the figure are target-defined and interpret the variables set in the \texttt{Ingress} control. If \texttt{m.drop} is set, the packet will be dropped by the traffic manager, skipping execution of the entire egress. If the \texttt{resubmit.emit()} is called, \texttt{m.recirculate} will implicitly be set, causing \toolname to reset all metadata and reroute the execution back to the ingress parser. We have modeled this control flow for targets such as \vonemodel, \tna, and \ttwona.

\subsection{Handling Target-Specific Behavior}
Targets have different intra-block semantics and diverge in their interpretation of core \pfour language constructs. \toolname is structured such that every function in the abstract machine can be overridden by target extensions. For example, the \vonemodel \toolname extension overrides the canonical \toolname table continuation to implement its own annotation semantics (\eg the ``priority'' annotation, which reorders the execution of constant table entries based on the value of the annotation). Targets may also reinterpret core parsing functions (\eg \texttt{extract}, \texttt{advance}, \texttt{lookahead}). 

\subsubsection{\toolname's Approach to Packet-Sizing}
\label{sec:design_packet_sizing}
One area where there is significant diversity among targets is in the semantics of operations that change the size of the packet. Some paths in a \pfour program are only executable with a specific packet size. \pfour externs such as \texttt{extract} can throw exceptions when the packet is too short or malformed. These packet paths are often sparsely tested when developing a new \pfour target and toolchain. Particularly on hardware targets, packets with an unexpected size may not be parsed as expected. Correspondingly, \toolname must be able to control the size of the input packet (Challenge \ref{para:challenge_2}). And, since some of these inputs may trigger parser exceptions, it also needs to model the impact these exceptions have on the content and length of the packet.

%%Anirudh: Resume here.
\toolname implements packet-sizing by making the packet size a symbolic variable in the set of path constraints. This encoding turns out to be non-trivial. Since the required packet size to traverse a given path is now a symbolic variable, it is only known after the SMT solver is invoked. However, at the same time, externs in \pfour manipulate the size of the packets (\eg \texttt{extract} calls shorten while \texttt{emit} calls lengthen the packet), which requires careful bookkeeping in first-order logic. Targets also react differently to specific packet sizes (\eg \bmv produces garbage values for 0-length packets~\cite{bmv2_outputs}, whereas Tofino drops packets smaller than 64 bytes~\cite[\S 7.2]{open_tofino}). Lastly, some targets add and remove content from the packet (\eg Tofino adds internal metadata to the packet~\cite[\S 5.1]{open_tofino}). Any packet-sizing mechanism needs to handle these challenges, while remaining target independent.

Our approach is to model packet-sizing as described in the \pfour specification. For each program path, we calculate the \emph{minimum} header size required to successfully exercise the path without triggering a parser exception. The packet-sizing model defines and manipulates three symbolic bit vector variables: the required input packet ($I$), the live packet ($L$), and the emit buffer ($E$). The input packet $I$ represents the \emph{minimum} header content required to reach a particular program point without triggering an exception. The live packet $L$ represents the packet header content available to the interpreter stepping through the \pfour program, \eg \texttt{extract} will consume content from $L$. The emit buffer $E$ is a helper variable which accumulates the headers produced by \texttt{emit}. This is necessary to preserve the correct order of headers, as prepending headers to $L$ each time \texttt{emit} is executed would cause it to be inverted.

Initially, all variables are zero-width bit vectors. While traversing the program, parser externs (\eg \texttt{extract} or \texttt{advance}) in the \pfour program slice data from the live packet $L$. If $L$ is empty (meaning we have run out of packet header data), \toolname allocates a new symbolic packet header and adds it to $I$. Targets may augment the input packet with custom parsable data (\eg metadata) which reduces the input packet needed to avoid triggering a parser exception. Correspondingly, this content is added to the live packet variable $L$. Once \toolname has finished executing a path, $I$ will denote the content of the final input packet in the generated test. $L$ on the other hand will correspond to the content of the expected packet output. Fig.~\ref{fig:packet_sizing} in App. \ref{app:packet_sizing} illustrates the variables used for an example pipeline.

This design also handles multi-parser, multi-pipe targets, such as Tofino. Each Tofino pipeline has two parsers: ingress and egress. The egress parser receives the packet ($L$) after the ingress and traffic manager. If the egress parser runs out of content in $L$, \toolname must again append symbolic content to $I$, increasing the size of the minimum packet required to parse successfully. 

\subsection{Controlling Unpredictable Behavior}
\label{sec:design_taint}

Many \pfour programs are non-deterministic, which can lead to unpredictable outputs (Challenge \ref{para:challenge_3}). To avoid generating ``flaky'' tests, we use taint analysis~\cite{taint}. As \toolname steps through the program, we keep track of which bits have a known value (i.e., ``untainted''), and which bits have an unknown value (i.e., ``tainted''). For example, a declaration of a variable that is not initialized and reads from random memory will be designated as tainted. The result of any operation that references a tainted variable will also be tainted. Later, when generating tests, we use the taint to avoid generating tests that might fail---\ie due to testing tainted values. For example, if the output packet contains taint, we know that certain bits are unreliable. We use test-framework-specific facilities (\eg ``don't care'' masks) to ignore tainted output bits. On the other hand, if the output port is tainted and the test framework does not support wildcards for the output port, \toolname can not reliably predict the output, so we drop the test and issue a warning.

\Para{Mitigating taint spread.}
A common issue with taint analysis is \emph{taint spread}, the proliferation of taint throughout the program, quickly tainting most of the state. In extreme situations, taint spread can make test generation almost useless, as the generated tests have many ``don't care'' wild cards. To mitigate taint spread we use a few heuristics. First, we apply optimizations to eliminate unnecessary tainting (for example, multiplying a tainted value with \texttt{0} results in \texttt{0}). Second, we exploit freedom in the \pfour specification to avoid taint. For example, when a ternary table key is tainted, we insert a wildcard entry that always matches. Third, we model target-specific determinism. For example, the Tofino compiler provides an annotation which initializes all target metadata with 0. Applying these heuristics significantly reduces taint in practice. 

\Para{Applying taint analysis.}
In our experience, taint analysis is essential for ensuring that \toolname can generate predictable tests. It substantially reduces the signal-to-noise ratio for validation engineers, enabling them focus on analyzing genuine bugs rather than debugging flaky tests. And, although it was not intended for this purpose, \toolname's taint analysis can be used to track down undefined behavior in a \pfour program.  \toolname does this by offering a ``restricted mode,'' which triggers an assertion when the interpreter reads from an undefined variable on a particular path. The more ``correct'' a \pfour program is written (\ie by carefully validating headers) the less taint (and fewer assertions) it produces.

\Para{Prototyping extensions using taint.}
Another useful byproduct of taint analysis is the ability to easily prototype a \toolname extension and its externs. Rather than implementing the entire \toolname extension at once, a developer can substitute taint variables for the parts that may need time-intensive development (a form of \emph{angelic programming}~\cite{angelic_nondeterminism}). By constraining the non-determinism of the unimplemented parts of the extension it is possible to generate deterministic tests early. We used this approach to generate initial stubs for many externs (e.g., checksums, meters, registers) before implementing them precisely. 

\subsection{Supporting Complex Functions}
\label{sec:design_concolic}
To handle complex functions that cannot be easily encoded into first-order logic (Challenge \ref{para:challenge_4}), \toolname uses \emph{concolic execution}~\cite{dart,cute}. Concolic execution is an advanced technique that combines symbolic and concrete execution. In a nutshell, it leaves hard-to-model functions unconstrained initially, and adds constraints later using the concrete implementation of the function. The \texttt{verify\_checksum} function described in \S~\ref{sec:examples} is an example where concolic execution is necessary. The checksum computation is too complex to be expressed in first-order logic. Instead, we model the return value of the checksum as an uninterpreted function dependent on the input arguments of the extern. While \toolname's interpreter steps through the program, this uninterpreted function acts as a placeholder. If the function becomes part of a path constraint, the SMT solver is free to fill it in with any value that satisfies the constraint.

Once we have generated a full path, we need to assign a concrete value to the result of the uninterpreted function. First, we invoke the SMT solver to provide us with concrete values to the input arguments of the uninterpreted function that satisfy the path constraints we have collected on the rest of the path. Second, we use these input arguments as inputs to the actual extern implementation (\eg the hash function executed by the target). Third, we add equations to the path constraints that bind all the values we have calculated to the appropriate input arguments and output of the function. We then invoke the solver a second time to assess whether the result computed by the concrete function satisfies all of the other constraints in the path. If so, we are done and can generate a test with all the values we calculated. 

\Para{Handling unsatisfiable concolic assignments.}
 In some cases, the newly generated constraints cannot be satisfied using the inputs chosen by the SMT solver. In practice, retrying by generating new inputs may not lead to a satisfiable outcome. Before discarding this path entirely, we try to apply function-specific optimizations to produce better constraints for the concolic calculation. For example, the \texttt{verify\_checksum} function (see also \S \ref{sec:examples}) tries to match the computed checksum of input data with an input reference value. If the computed checksum does not match with the reference value, \texttt{verify\_checksum} reports a checksum mismatch. Instead of retrying to find a potential match, we add a new path that forces the reference value to be equal to the computed checksum. This path is satisfiable if the reference value is derived from symbolic inputs, which is often the case. Note that in situations where the reference value is a constant, we are unable to apply this optimization.

%-------------------------------------------------------------------------------
\section{Path Selection Strategies}
%-------------------------------------------------------------------------------
\label{sec:useful_tests}
Methodologies that assess the program coverage of tests have become standard software engineering practice. While path coverage is often infeasible (as the number of paths grows exponentially), statement coverage, also known as line coverage, has been proposed as a good metric for evaluating a test suite~\cite{klee}. \toolname allows users to pick from several different path selection strategies to produce more diverse tests, including \randcov and \maxcov. As the name suggests, \randcov simply jumps back to a random known branch point in the program once \toolname has generated a test. \maxcov is similar to the concept with the same name in Klee~\cite{klee}. After a new test has been generated, it selects the first path from all unexplored paths it has seen so far which will execute \pfour statements that have not been covered. If no path with new statements can be found, \maxcov falls back to random backtracking until a path with new statements is discovered. This greedy search covers new statements quickly, but at the cost of higher memory usage (because it accumulates unexplored paths with low potential) and slower per-test-case performance. We measure in \S \ref{sec:evaluation_coverage} how these strategies perform on large \pfour programs. Our path selection framework is extensible, allowing us to integrate many different selection strategies. We can easily add other success metrics, such as table, action, or parser state coverage.

\Para{Targeted test generation with preconditions.}
Path selection strategies guide test case generation towards a goal, but they do not select for a specific type of test. \toolname also gives users the ability to instrument their \pfour program with a custom extern (\texttt{testgen\_assume}). This \toolname-intrinsic extern adds a path constraint on variables accessible within the \pfour program (\eg \texttt{h.eth\-\_hdr.eth\-\_type} == \texttt{0x0800}), which forces \toolname to only produce tests that satisfy the provided constraint. Assume statements are similar to p4v's assumptions~\cite{p4v}, Vera's NetCTL constraint~\cite{vera}, or Aquila's LPI preconditions~\cite{aquila}. We study the effect of these constraints in \S \ref{sec:evaluation_coverage}.

\Para{Instrumenting fixed control-plane configurations.}
Network operators in general have restricted environments in which only a limited set of packets and control plane configuration is actually valid. Similar to Meissa~\cite{meissa} and SwitchV~\cite{switchv}, we are developing techniques to instrument a particular fixed control plane configuration before generating tests. We are looking into a specification method to allows users to only generate tests which comply with their environment assumptions. As an initial step in this direction, \toolname implements SwitchV's \pfourconstraints framework (\S \ref{sec:implementation_v1model}).

%-------------------------------------------------------------------------------
\section{Implementation}
%-------------------------------------------------------------------------------
\label{sec:implementation}
\toolname is written as an extension to \pfourc using about 19k lines of C++ code, including both \toolname core and its extensions. To resolve path constraints, \toolname uses the Z3~\cite{z3} SMT solver.

% Different levels of precision can lead to a highly varying number of branches being generated. A rather simple \texttt{max} extern, which returns the larger of two values, can be implemented by simply picking one of the values and adding a constraint that it needs to be larger. The function can also be modeled to branch into three different paths instead. One where the input values are equal, one where the first value is larger, and one where the second value is larger.

\Para{Interacting with the control plane.}
\toolname uses the control plane to trigger some paths in a \pfour program (\eg paths dependent on parser value sets~\cite[\S 12.11]{p416_spec}, tables, or register values). Since \toolname does not perform load or timing tests, the interaction with the control plane is mostly straightforward. For each test that requires control-plane configuration, \toolname creates an abstract test object, which becomes part of the final test specification. For tables, \toolname creates forwarding entries, and if the test framework provides support, it can also initialize externs such as registers, meters, counters and check their final state after execution. In general, richer test framework APIs give \toolname more control over the target---\eg STF lacks support for range-based match types, which means some paths cannot be executed. 

\subsection{\toolname Extensions}
\label{sec:implementation_ext}
%---- FIGURE ----%
%%%%%%%%%%%%%%%%%%%%%%%%%%%%%%%%%%%%%%%%%%%%%%%%%%%%%%%%% Targets and test back ends
 \begin{table}[!t]
 \centering
 \begin{footnotesize}
 \begin{tabular}{@{}llll@{}}\toprule
 Architecture & Target & Test back end  & C/C++ LoC\\ \midrule
\texttt{v1model} & \bmv & STF, PTF, Protobuf, Meta & 5289  \\ 
\texttt{tna} & Tofino 1 & Internal, PTF  & 475 (3314 shared) \\
\texttt{t2na} & Tofino 2 & Internal, PTF & 478 (3314 shared) \\
\texttt{ebpf\_model} & Linux Kernel & STF & 1011 \\
\texttt{pna} & DPDK SoftNIC & PTF, Meta & 1694 \\
\bottomrule
\end{tabular}
\caption{\toolname extensions. The core of \toolname is 6679 LoC.}
\label{tbl:target_impl}
\end{footnotesize}
\end{table}
%---- FIGURE ----%
Tbl.~\ref{tbl:target_impl} lists the targets we have instantiated with \toolname. We also list the LoC every extension required, noting that \tna and \ttwona share a lot of code. Further, \vonemodel LoC are inflated because of the \pfourconstraints parser and lexer implementation specific to the \vonemodel extension. We modeled the majority of the Tofino externs based on the \pfour Tofino Native Architecture (TNA) available in the Open-Tofino repository~\cite{open_tofino}. Each extension also contains support for several test frameworks. The \vonemodel instance supports PTF, STF, Protobuf~\cite{protobuf} messages, and the serialization of metadata state. The Tofino instance supports PTF and an internal compiler testing framework. The eBPF instance supports STF. The Portable NIC Architecture (PNA)~\cite{pna_spec} instance only has metadata serialization.

\subsubsection{\vonemodel}
\label{sec:implementation_v1model}

\toolname supports the \vonemodel architecture, including externs such as \texttt{recirculate}, \texttt{verify\_checksum}, and \texttt{clone}. The \texttt{clone} extern requires \toolname's entire toolbox to model its behavior, so we explain it in detail below.

\Para{Implementing clone.}
The \texttt{clone} extern duplicates the current packet and submits the cloned packet into the egress block of the \vonemodel target. It alters subsequent control flow based on the place of execution (ingress vs. egress control). Depending on whether \texttt{clone} was called in the ingress vs. egress control, the content of the recirculated packet will differ. Further, which user metadata is preserved in the target depends on input arguments to the \texttt{clone} extern.

We modeled this behavior entirely within the \bmv extension to \toolname without having to modify the core code of \toolname's symbolic executor. We use the pipeline control flow and continuations to describe \texttt{clone}'s semantics, concolic execution to compute the appropriate clone session IDs, and taint tracking to guard against unpredictable inputs.

\Para{\pfourconstraints.}
\toolname's \bmv extension also implements the \pfourconstraints framework~\cite{switchv} for \vonemodel. \pfourconstraints annotates tables to describe which control plane entries are valid for this table. \pfourconstraints are needed for programs such as \middleblock~\cite{middleblock}, which models an aggregation switch in Google's Jupiter network~\cite{jupiter} that only handles specific entries. To generate valid tests for such programs, \toolname must accommodate constraints on entries. It does so by converting \pfourconstraints annotations into its own internal predicates, which are applied as preconditions, restricting the possible entries, and hence, the number of generated tests (\S\ref{sec:evaluation}).

\subsubsection{\tna/\ttwona}
\label{sec:implementation_tna}

We have implemented the majority of externs for \tna and \ttwona, including meters, checksums, and hashes. For others, such as registers, we make use of rapid prototyping using taint. Our \ttwona extension leverages much of the \tna extension, but \ttwona is richer, so it took more effort to model its capabilities. Not not only does \ttwona use different metadata, it also adds a new programmable block (``ghost'') and doubles the number of externs. Also, both \tna and \ttwona support parsing packets at line-rate, which is significantly more complex than \bmv~\cite[\S 5]{open_tofino}.

\Para{Parsing packets with Tofino.}
The Tofino targets prepend multiple bytes of metadata to the packet~\cite[\S 5.1]{open_tofino}. As an Ethernet device, they also append a 32-bit frame check sequence (FCS) for each packet. Both the metadata and FCS can be extracted by the parser but are not part of the egress packet in the emit stage. If the packet is too short and externs in the parser trigger an exception, Tofino drops the packet in the ingress parser, but not in the egress parser~\cite[\S 5.2.1]{open_tofino}. However, if the ingress control does read from the \texttt{parser\_error} metadata variable, the packet is not dropped and instead skips the remaining parser execution and advances to the ingress control. The content of the header that triggered the exception is unspecified in this case. We model this behavior entirely in the Tofino instantiations of \toolname. We treat the metadata, padded content, and FCS as taint variables which are prepended to the live packet $L$. Since Tofino's parsing behaves differently to the description in the \pfour specification, we extend the implementations of \texttt{advance}, \texttt{extract}, and \texttt{lookahead} in the Tofino extensions to model the target-specific behavior.

\subsubsection{\ebpfmodel}
\label{sec:implementation_ebpf}
As a proof of concept for \toolname's extensibility we also implemented an extension for an end-host target. \ebpfmodel is a fairly simple target, but it differs from \tna and \ttwona, which are switch-based. The pipeline has a single parser and control. The control is applied as a filter following the parser. There is no deparser. The eBPF kernel target rejects a packet based on the value of the \texttt{accept} parameter in the filter block. If \texttt{false}, the packet is dropped. As there is no deparser, we model implicit deparsing logic by implementing a helper function that iterates over all headers in the packet header structure and emits headers based on their validity. We were able to implement the eBPF target in a few hours and generate \inout tests for all the available programs (30) in the \pfourc repository. Because of the lack of maturity of the target, we did not track any bugs in the toolchain. 

\subsubsection{\pna}
\label{sec:implementation_pna}
PNA~\cite{pna_spec} is a P4 architecture describing the functionality of end-host networking devices such as (Smart-)NICs. A variety of targets using the \pna architecture have been put forward by Xilinx~\cite{xilinx}, Keysight~\cite{p4_keysight_workshop}, NVIDIA~\cite{nvidia_doca_sdk}, AMD~\cite{pensando_p4}, and Intel~\cite{ipu}. We have instantiated a \toolname extension for a publicly available \texttt{pna} instance, the DPDK SoftNIC~\cite{dpdk_softnic}. Since there are no functional testing frameworks (\eg PTF or STF) yet available for this target, we generate abstract test templates, which describe the \inout behavior and expected metadata after each test. By generating these abstract tests we can already perform preliminary analysis on existing \pna programs (\S\ref{sec:evaluation_coverage}).
%-------------------------------------------------------------------------------
\section{Evaluation}
\label{sec:evaluation}
Our evaluation of \toolname considers several factors: performance, correctness, coverage, and effectiveness at finding bugs.

%---- FIGURE ----%
%%%%%%%%%%%%%%%%%%%%%%%%%%%%%%%%%%%%%%%%%%%%%%%%%%%%%%%%% Performance
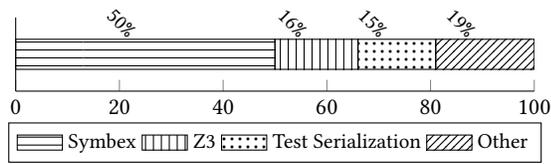
\begin{figure}[t]
\centering
\pgfplotsset{perfbar/.style={
xbar stacked,
area style,
width=\columnwidth,
xmin=0,xmax=100,
ymin=-0.7,ymax=0.7,
bar width=4mm, y=7mm,
nodes near coords={\pgfmathprintnumber\pgfplotspointmeta\%},
every node near coord/.append style={rotate=-45,font=\fontsize{8}{8}\ttfamily, left,shift={(axis direction cs:-3.1,0.1)}},
% point meta=x
% axis line style={draw=none}, % don't draw the axis lines
ytick=\empty,
% x axis line style={stealth},
% enlarge x limits=false,
% y axis line style={stealth},
% xtick pos=bottom,
axis y line*=none,
axis x line*=none,
}}
\begin{tikzpicture}
\begin{axis}[perfbar,legend style={area legend, at={(xticklabel cs:0.5)}, anchor=north, legend columns=-1, font=\small}] 
    \addplot[draw=black, fill=red!40, pattern=horizontal lines] coordinates{(50,0)};
    \addplot[fill=gray!40, pattern=vertical lines] coordinates{(16,0)};
    \addplot[fill=green!40, pattern=dots] coordinates{(15,0) };
    \addplot[fill=orange!40, pattern=north east lines] coordinates{(19,0)};
   \legend{Symbex,Z3,Test Serialization, Other}
\end{axis}
\end{tikzpicture}

% \begin{tikzpicture}
% \pie[scale=0.42, text=inside, rotate=180, sum = 100, color={black!10, black!20, black!30, black!40}, font=\footnotesize]
% {   50/Symbolic\\Executor,
%     16/Z3,
%     15/Test\\ser.,
%     19/Other
% }
% \end{tikzpicture}

\caption{Average CPU time spent in \toolname.}
\label{fig:performance}
\end{figure}

%---- FIGURE ----%
%-------------------------------------------------------------------------------
\subsection{Performance}
\label{sec:evaluation_performance}
To evaluate \toolname's performance when generating tests, we measured the percentage of cumulative time spent in three major segments: 1) stepping through the symbolic executor, 2) solving Z3 queries, 3) serializing an abstract test into a concrete test. Fig.~\ref{fig:performance} shows \toolname's CPU time distribution for generating 10000 tests for the larger programs listed in Tbl.~\ref{tbl:prog_stats}. In general, solving path constraints in Z3 accounts for around 16\% of the overall CPU time. \toolname spends the majority of time in the symbolic executor. This is expected, as we prioritized extensibility and debuggability for \toolname's symbolic execution engine, not performance. We expect performance to improve as the tool matures. From informal conversations we are aware that \toolname generates tests on the same order of efficiency as SwitchV's \texttt{p4-symbolic} tool does.

\subsection{Correctness}
\label{sec:evaluation_completeness}
As a general test-oracle, \toolname is designed to support multiple targets. We consider our design successful if a target extension is both able to generate correct test files for a wide variety of \pfour programs and also produce tests that pass for complex, representative programs on each target.

\Para{Producing valid tests for diverse \pfour programs.}
\label{sec:evaluation_correct} To ensure that \toolname's interpretations of \pfour and target semantics are correct, we generated tests for a suite of programs and executed them on the target. For \vonemodel, \texttt{pna}, and \texttt{ebpf\_model}, we selected all the \pfour programs available in the \pfourc test suite. For Tofino, we used the programs available in the P4Studio SDE and a selected set of compiler tests given to us by the Tofino compiler team. The majority of these programs are small and easy to debug, as they are intended to test the Tofino compiler. In total, we tested on 458 Tofino, 191 Tofino 2, 507 \bmv, 62 PNA, and 30 eBPF programs. 

We used \toolname to generate 10 \inout tests with a fixed random seed for each of the above programs. We then executed these tests using the appropriate software model and test back ends. In fact, on every repository commit of \toolname, we execute \toolname on all \numextensions extensions and their test back ends (Tbl.~\ref{tbl:target_impl}), totaling more than 2800 \pfour programs and 10 tests per program.  We used this technique to progressively sharpen our semantics over the course of a year, running \toolname millions of times. If the execution of a test did not lead to the output expected by \toolname, we investigated. Sometimes,  it  was a bug in \toolname, which we fixed. Sometimes, the target was at fault and we filed a bug (see \S \ref{sec:evaluation_utility}). 

\Para{Producing valid tests for large \pfour programs.}
For the \vonemodel, we chose two actively maintained \pfour models of real-world data planes. \middleblock (\S~\ref{sec:implementation_v1model}) and \upfour~\cite{up4}. \upfour is a \pfour program developed by the Open Networking Foundation (ONF) which models the data plane of 5G networks. We have considered other programs but they were either written in P4\textsubscript{14}~\cite{switchp4} or not sufficiently complex to provide a useful evaluation~\cite{princeton_cabernet}. For \tna/\ttwona, we generate tests for the appropriate version of \switchpfour, the most commonly used \pfour program for the Tofino programmable switch ASIC. We execute the generated tests on either \bmv or the Tofino model (a semantically accurate software model of the Tofino chip). For each target, we generate 100 PTF tests. The eBPF kernel target does not have a suite of representative programs. Instead, we generated tests for \pfourc's sample programs. The tests we have generated pass, showing that we can correctly generate tests for large programs. \pna on the DPDK SoftNIC does not have an end-to-end testing pipeline available yet, but we still generate tests for its programs. As a representative program we picked \dash, which models the end-to-end behavior of a programmable data plane in \pfour~\cite{dash_p4_model}. \dash is still under development, but is already complex enough to generate well over a million unique tests.

\subsection{Coverage}
\label{sec:evaluation_coverage}

%%%%%%%%%%%%%%%%%%%%%%%%%%%%%%%%%%%%%%%%%%%%%%%%%%%%%%%%% Coverage statistics 
\begin{table}[t]
\begin{footnotesize}
\begin{tabular}{@{}lllll@{}}
\toprule
\pfour program & Valid tests & Time & Stmts. & Stmts. covered \\ \midrule
\middleblock \texttt{(v1model)} & 74472 & \textasciitilde40m & 150 & 100\% \\
\upfour \texttt{(v1model)} & 57853 & \textasciitilde55m & 185 & 100\%   \\
\dash \texttt{(pna)} & >1M & \textasciitilde668m & 256 & \textasciitilde90\%   \\
\tnasimpleswitch \texttt{(tna)} & >1M & \textasciitilde628m & 300 & \textasciitilde43\%   \\
\switchpfour \texttt{(tna)} & >1M & \textasciitilde2653m & 921 & \textasciitilde36\%  \\
\switchpfour \texttt{(t2na)} & >1M & \textasciitilde2719m & 1024 & \textasciitilde31\% \\
\bottomrule
\end{tabular}
\caption{Coverage statistics for large \pfour programs using DFS.}
\label{tbl:prog_stats}
\end{footnotesize}
\end{table}

When generating tests, \toolname tracks the statements (after dead-code elimination) covered by each test. Once \toolname has finished generating tests, it emits a report that details the total percentage of statements covered. We use this data to identify any \pfour program features that were not exercised. For example, some program paths may only be executable if the packet is recirculated. 

\Para{How well does \toolname cover large programs?} 
We tried to exhaustively generate tests for the programs chosen in the previous section. Tbl.~\ref{tbl:prog_stats} provides an overview of the number of tests generated for each program (this number correlates with the number of possible branches as modelled by \toolname) and the best statement coverage we have achieved using \dfscov. As expected, for the \switchpfour programs of \tna and \ttwona, we generate too many paths to terminate in a reasonable amount of time. For the \switchpfour programs we list the coverage we achieved before halting generation after the millionth test.

\Para{How does path selection help with statement coverage? }
Tbl.~\ref{tbl:prog_stats} shows that the number of tests generated for larger \pfour programs can be overwhelming. In practice, users want tests with specific properties, which necessitates the use of path selection strategies. We measure the effect of the \toolname's path selection strategies (\S \ref{sec:useful_tests}). We select \middleblock and \upfour as representative sample programs for \vonemodel. For \tna and \ttwona, we select \tnasimpleswitch, which we patched up such that all statements in the program are reachable. We have chosen \tnasimpleswitch for two reasons: (i) we have not implemented all features to fully cover \switchpfour (specific register/meter configurations, recirculation) to achieve full statement coverage,\footnote{We currently achieve around 90\% coverage using \maxcov.} and (ii) \tnasimpleswitch is an open-source program available at the OpenTofino repository~\cite{open_tofino}. \tnasimpleswitch is still a complex Tofino program: it produces over 30 million unique, valid tests. We generate tests with each strategy until we hit 100\% statement coverage. We compare \randcov and our \maxcov to standard \dfscov. We measure the total number of tests needed to achieve coverage across a sample of 10 different seeds.

\begin{figure}[t]
    \centering
    \includegraphics[width=\linewidth]{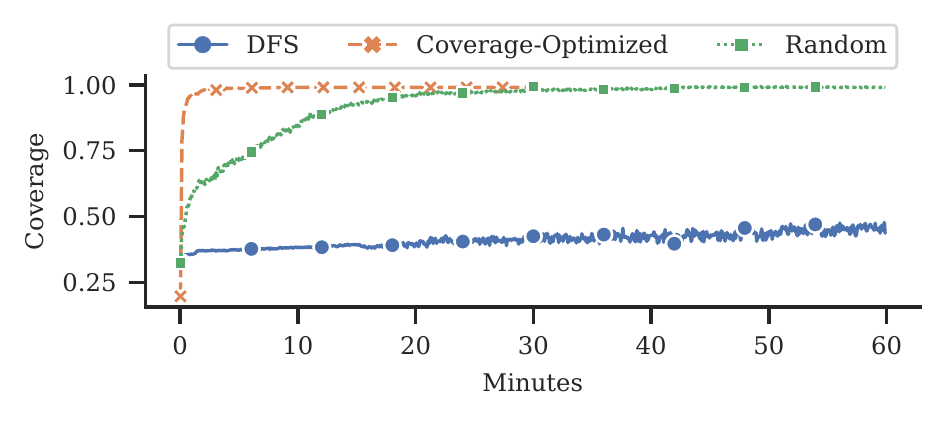}
    % \vspace{-2.5em}
    \caption{Path selection strategy performance on \tnasimpleswitch.}
    \label{fig:strategy_coverage}
\end{figure}

Fig.~\ref{fig:strategy_coverage} shows the mean coverage across 10 seeds over 1000 timesteps for \tnasimpleswitch. We stopped a heuristic if it did not achieve 100\% within an hour of generating tests. Only \maxcov reliably accomplishes full coverage in this time frame and outperforms \randcov and \dfscov by a wide margin. \maxcov always outperforms \dfscov and generally outperforms \randcov. In some cases, however, (\eg \upfour) \maxcov is not sophisticated enough to find the path which covers a sequence of statements. In those cases, it will perform similarly to \randcov.  Tbl.~\ref{tbl:strategies} of the appendix shows a results breakdown for all selected programs.

\Para{How do preconditions affect the number of generated tests?}
We conducted a small experiment to measure the impact of applying preconditions and simplified extern semantics on \middleblock. We measured the number of generated tests when fixing the input packet size (thus avoiding parser rejects in externs) and applying SwitchV's \pfourconstraints. Fig.~\ref{fig:precondition_reduction} shows the results. The number of generated tests can vary widely, based on these input parameters. Applying the input packet size and the \pfourconstraints table entry restrictions can reduce the number of generated tests by as much as 71\%. Adding \texttt{testgen\_assume} (\S \ref{sec:useful_tests}) statements, which mandates that we only produce packets with TCP/IP headers, reduces the generated tests by 95\%. Tbl.~\ref{tbl:middleblock_reduction} in the appendix has detailed statistics.

\Para{What are the limits of \toolname's statement coverage?} There are \pfour programs where \toolname can not achieve full statement coverage. An example is \blinkpfour~\cite{blink}, a \pfour program, where statement execution depends on the timestamp metadata field which is set by the target when a packet is received. Since \toolname can not control the initialization of the timestamp for \bmv (yet), we are unable to cover any statement depending on it. Other tools such as FP4~\cite{fp4} and P4wn~\cite{p4wn} are able to cover these statements as they generate packet sequences which may eventually cause the right timestamp to be generated. 
This limitation is not insurmountable. In the future, we plan to mock timestamps using a match-action table, or add an API for controlling timestamps directly.

%%%%%%%%%%%%%%%%%%%%%%%%%%%%%%%%%%%%%%%%%%%%%%%%%%%%%%%%% Middleblock Tests 

\subsection{\toolname in Practice}
\label{sec:evaluation_utility}
We have used \toolname to successfully generate tests for nearly a year. Compiler developers rely on \toolname to gain confidence in the implementation of new compiler features. For instance, they can generate tests for an existing program, enable the new compiler feature, and check that the tests still pass. This approach identified several flaws in new compiler targets and features during development. We have also used \toolname to give users of Tofino confidence to upgrade their targets or their toolchains. In one of our use cases, a switch vendor had reservations on migrating their \pfour programs from Tofino 1 to Tofino 2. The vendor could not ensure that the behavior of the program remained semantically equivalent in this new environment. Using \toolname we generated a high-coverage test suite, which reassured the team that they could safely migrate to the Tofino 2 chip. 

\Para{Generating tests for abstract network device models.}
An increasingly popular use-case of \pfour is to use it as a modeling language to describe network data planes~\cite{switchv,dash_p4_model}. Often, these data plane models lack tests. \toolname can exhaustively generate tests for the \pfour data plane model, where the tests also satisfy particular coverage criteria. Further, because \toolname is extensible, a developer modelling their device can use arbitrary \pfour architectures. We are now working with the DASH~\cite{dash_p4_model} and SwitchV~\cite{switchv} developer teams, who are interested in applying \toolname to their data plane models written for the \texttt{pna} and \texttt{v1model} architectures.

\subsubsection{Bugs}
\label{sec:evaluation_bugs}

For any validation tool, the bottom line is whether it effectively finds bugs, particularly in mature, well-tested systems. To evaluate \toolname's effectiveness, we used the workflow described in \S\ref{sec:evaluation_correct}, by running \toolname on each program in the appropriate test suite. Tbl.~\ref{tbl:bug_report} summarizes the bugs we found. Tbl.~\ref{tbl:bug_list} in the appendix provides details on the bugs we have filed for \bmv. For confidentiality reasons, we are unable to provide details on Tofino bugs.

\begin{figure}[t]
    \centering
    \includegraphics[width=\linewidth]{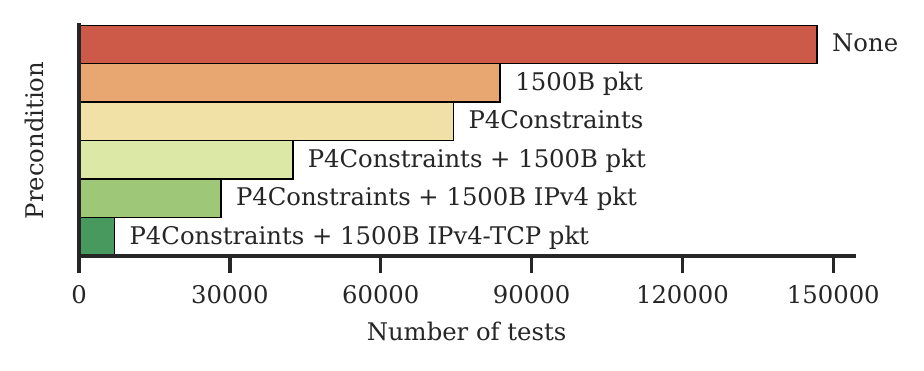}
    % \vspace{-2.5em}
    \caption{Effects of preconditions on the total number of tests generated for \middleblock.}
    \label{fig:precondition_reduction}
\end{figure}

\begin{table}[t]
\centering
\begin{footnotesize}
\begin{tabular}{l|lllr}
\hline
\textbf{Bug Type}           & Feature                   & \bmv                   & Tofino                                & \textbf{Total}                           \\ \hline
\multirow{5}{*}{Exception}  & Unusual path          &                  2                     &                       6                & 8                                \\ 
                            & Synthesized control plane &                  5                     &                   1                    & 6                                                             \\
                            & Packet-sizing             &                  0                     &                   2                     & 2                                                             \\
                            & Extern model              &                  0                     &                   1                    &   1                                                            \\ \cline{2-5}
                            & Total                     & \bmvbugsexcept         & \tnabugsexcept         & \multicolumn{1}{r}{\bugsexcepttotal} \\ \hline
\multirow{5}{*}{Wrong Code} & Unusual path         &                    0                   &                    5                   & 5                              \\
                            & Synthesized control plane &                   1                    &                1                       &      2                                                        \\
                            & Packet-sizing             &                   0                    &              0                        &     0                                                         \\
                            & Extern model              &                   0                    &               1                        &     1                                                         \\ \cline{2-5}
                            &  Total                         & \bmvbugswrong          & \tnabugswrong          & \bugswrongtotal  \\ \hline
\multicolumn{2}{c}{\textbf{Total}}                         & \textbf{\bmvbugstotal} & \textbf{\tnabugstotal} & \textbf{\bugstotal}       \\ \hline
\end{tabular}
\caption{Bugs in targets discovered by \toolname.}
\label{tbl:bug_report}
\end{footnotesize}
\end{table}

\Para{What are the bugs we are interested in?}
We report only \emph{target stack bugs}---\ie a bug in the software or hardware stack. We consider a target stack bug any failing test that was generated by \toolname but was not an issue with \toolname itself. This includes compiler bugs as well as crashes of the control-plane software, driver, or software simulator. We only count bugs that are both new, distinct (\ie cause a new entry in the issue tracker), and non-trivial (bugs which require either a particular packet size, control-plane configuration, or extern to be exercised). If a bug is considered a duplicate by the developers we only count it once. \toolname revealed two types of bugs: (1) exceptions, where the combination of inputs caused an exception or other fault; and (2) wrong code bugs, where the test inputs did not produce the expected output. 

\Para{What caused these bugs?}
The causes of the bugs found were diverse. Some were due to errors in the compiler back end, others due to mistakes in the software model, while still others due to errors in the control plane software and test framework. For each bug, we filed an issue in the respective tracker system. Several issues either anticipated a customer bug that was filed later or reproduced an existing issue that was still open. In several instances, \toolname was able to discover bugs where hand-written tests lacked coverage.

\Para{What features of \toolname were important for finding a bug?}
8 of the total \bugstotal we have found were triggered by \toolname synthesizing table and extern configurations. 2 were triggered by \toolname implementing a detailed model of extern functions. 2 were triggered by \toolname generating tests with unexpected packet sizes.  The remaining bugs were caused because \toolname's generated tests exercised untested program paths or esoteric language constructs (\eg a stack-out-of-bounds error or header union access). Overall, we found more incorrect behavior bugs with Tofino because of (i) its complexity and (ii) the fact that we focused our bug-tracking efforts on Tofino and gave \bmv issues lower priority. 

\Para{Reachability bugs in \pfour programs}
A side-effect of \toolname's support of explicit coverage heuristics is its ability to detect reachability bugs in \pfour programs. In some cases, \greedycov is unable to cover a particular program statement. This may be because of failures in the heuristic, but often the code is simply non-executable---\ie dead.  We encountered several instances of such dead code for proprietary and public production-grade programs~\cite{dash_pipeline_dead_code}. The developers were usually appreciative of our bug reports, which occurred in complex programs that are difficult to debug, especially early in the development process.

%-------------------------------------------------------------------------------
\section{Related Work}
%-------------------------------------------------------------------------------
\begin{table*}[t]
\begin{footnotesize}
\centering
\begin{tabularx}{\linewidth}{llllll}
\hline
\textbf{Tool} & Input generation method & Synthesizes control-plane? & Multi-target? & Models target semantics? & Data plane coverage metric  \\
\hline
% ATPG~\cite{atpg} & Symbex & $\times$ & \checkmark & \checkmark \\
Meissa~\cite{meissa} & Symbolic Execution & $\times$ &  $\times$ & $\checkmark$ & Symbolic model \\
SwitchV (via \texttt{p4-symbolic})~\cite{switchv} & Symbolic Execution & $\times$  &  $\times$ & $\checkmark$ & Symbolic model, Assertions \\
\texttt{p4pktgen}~\cite{p4pktgen} & Symbolic Execution &  $\checkmark$ & $\times$ & $\times$ & Symbolic model \\
Gauntlet (model-based testing)~\cite{gauntlet} & Symbolic Execution &  $\times$ & $\checkmark$ & $\times$ & Symbolic model \\
% Petr4~\cite{petr4} & Differential testing & $\times$ & $\checkmark$ & $\checkmark$ & Symbolic model \\
PTA (uses \texttt{p4v})~\cite{pta} & Fuzzing &  $\times$  & $\checkmark$ & $\times$ & Symbolic model (\texttt{p4v}) \\
DBVal~\cite{dbval} & Fuzzing &  $\times$  & $\checkmark$ & $\times$ & Tables, Actions \\
FP4~\cite{fp4} & Fuzzing & $\times$ & $\checkmark$ & $\times$ & Actions \\
P6~\cite{fp4} & Fuzzing & $\times$ & $\checkmark$ & $\times$ & Symbolic model \\
\hline
\toolname  & Symbolic Execution &  $\checkmark$ & $\checkmark$ & $\checkmark$ & Symbolic model, source code \\
\hline
\end{tabularx}
\caption{\pfour tools generating \inout tests. Data plane coverage describes how the tool measures coverage of the generated inputs.}
\label{tbl:related_work}
\end{footnotesize}
\end{table*}
\label{sec:related_work}

% \Para{Network device testing.}
% In general, network device testing focuses on validating performance with respect to throughput, latency, and frame loss. For example, ITU-Y.1564~\cite{itu_y1564}, RFC 2544~\cite{rfc_2544}, and RFC 2889~\cite{rfc_2889} outline how to test Ethernet and LAN devices. Correctness is enforced in the form of validation of checksums, packet loss, or individual protocol fields (\eg the MAC destination must be preserved). Such standards are defined for each protocol layer or combination. Other work focuses on conformance testing, \ie validation that a particular protocol stack~\cite{protocol_validation}. \toolname is not designed to compete with this form of device validation and instead exploits \pfour's packet-processing description to produce protocol-independent, single-packet test centered around coverage.

\Para{Automatic test generation for Software-Defined Networks (SDNs).}
The SDN literature has considerable research dedicated to automated network testing, frequently using symbolic execution to verify the correctness of network invariants~\cite{anteater, sefl, nice, hsa, veriflow}. Some of these projects verify network data-plane configurations by generating test input packets, for example Automatic Test Packet Generation (ATPG)~\cite{atpg}. ATPG automates input packet generation to validate a configured switch network by computing all possible packets that cover every switch link and table rule. Monocle~\cite{monocle} and Pronto~\cite{pronto} are similar systems. All use the control-plane configuration as ground truth, which allows them to check whether the right packet headers have been forwarded out on the correct port. \toolname targets a richer data plane model than these prior approaches because the data plane is effectively specified in a DSL. But, \toolname focuses more narrowly on a single device's data plane implementation, not the entire network's forwarding rules.

\Para{Verifying \pfour programs.}
Many tools help verify \pfour programs against a formal specification. Tools in this domain usually rely on assertions that model relational properties---\eg the program does not read or write invalid headers~\cite{vera, p6, dataplane-equivalence, p4v, p4-assert, bf4, aquila, p4rl, p4wn}. \toolname is orthogonal to these tools. It produces tests for a \pfour program but does not check the correctness of the program itself.

Some of these tools~\cite{p6, p4v, p4rl, p4wn} are able to generate concrete test inputs in the form of input packets. The outputs of these inputs are then compared against developer-supplied assertions. In theory, with good assertions, this method can also detect bugs in a given \pfour toolchain. P6~\cite{p6} in particular considers ``platform-dependent bugs'', which are comparable to toolchain bugs.

\Para{Testing \pfour toolchains.}
Other tools focus on validating \pfour implementations by generating test inputs. Tbl.~\ref{tbl:related_work} provides a summary. Compared to \toolname, these tools are typically tailored to a single target or use case. \toolname relies on formal semantics to compute inputs and outputs, avoiding running a second system to produce the output~\cite{switchv, fp4}. In particular, developers using \toolname do not need to understand the semantics of the \pfour program to generate tests; \toolname provides these semantics as part of its tool.

\texttt{p4pktgen}~\cite{p4pktgen} is a symbolic executor that automatically generates tests. It focuses on the \texttt{v1model}, STF tests, and \bmv. In spirit, \texttt{p4pktgen} is close in functionality to \toolname. However, the tool does not implement all aspects of the \pfour language and \texttt{v1model} architecture---its capabilities as a test oracle are limited. We tried to reproduce the bugs listed in Tbl. \ref{tbl:bug_list} using \texttt{p4pktgen} but were not able to. \texttt{p4pktgen} either was not able to produce tests for the program or did not achieve the necessary coverage. While \texttt{p4pktgen} does support a form of packet-sizing to trigger parser exceptions, its model only considers a simple parser-control setup, not multiple subsequent parsers such as Tofino's.

SwitchV~\cite{switchv} uses differential testing to find bugs in switch software. It automatically derives inputs from a switch specification in \pfour, feeds the derived inputs into both the switch and a software reference model, and compares the outputs. SwitchV uses fuzzing and symbolic execution to generate inputs that cover a wide range of execution paths. To limit the range of possible inputs, the tool relies on pre-defined table rules and the \pfourconstraints framework. It also does not generate control-plane entries. Like \texttt{p4pktgen}, SwitchV is specialized to \texttt{v1model} and \bmv.

Meissa~\cite{meissa} is a symbolic executor specialized to the Tofino target. Meissa builds on the LPI language's pre- and post-conditions~\cite{aquila} to generate \inout tests. The tool is designed for scalability and uses techniques such as fixed match-action table rules, code summaries for multi-pipeline programs, and path pruning to eliminate invalid paths according to the input specification. \toolname's preconditions and path selection strategies combat the same scaling issues as Meissa. Meissa's source code is proprietary, which precludes a direct comparison.

PTA~\cite{pta} and DBVal~\cite{dbval} both implement a target-inde\-pen\-dent test framework  designed to uncover bugs in the \pfour toolchain. Both PTA and DBVal augment the \pfour program under test with extra assertions to validate the correct execution of the pipeline at runtime. Both projects provide only limited support for test-case generation. 

FP4~\cite{fp4} is a target-independent fuzzing tool that uses a second switch as a fuzzer to test the implementation of a \pfour program. FP4 automatically generates the necessary table rules and input packets to cover program paths. To validate whether outputs are correct, FP4 requires custom annotations instrumented by the user.

\Para{Coverage.}
There are important differences in testing tools assess coverage---see Tbl.~\ref{tbl:related_work} for a summary. \toolname marks a node in the source \pfour program as covered when the symbolic executor steps through that node and generates a test. FP4 measures action coverage by marking bits in the test packet header to track which actions were executed. As FP4 generates packets at line rate it achieves coverage for actions faster than \toolname. \texttt{p4pktgen} discusses branch coverage, which can be estimated by parsing generated tests to see which control-plane constructs (tables, actions) were executed. Meissa reports coverage based on the branches of its own formal model of the \pfour program. SwitchV also measures branch coverage based on developer-provided goals derived from its symbolic model. Another important consideration is whether programmers can annotate the program with constraints or preconditions---see Fig~\ref{fig:precondition_reduction}. In many scenarios, these constraints are necessary to model assumptions made by the overall system, but they also affect coverage since they reduce the number of legal paths.

\Para{Extensibility.}
Petr4~\cite{petr4} and Gauntlet~\cite{gauntlet} are designed to support multiple \pfour targets. Petr4 provides an ``plugin'' model that allows the addition of target-specific semantics. However, it does not support automatic test case generation and does not aim to provide path coverage. Gauntlet can generate \inout tests for multiple \pfour targets but it does not model externs, nor does it implement whole-program semantics to model the tested target.

%-------------------------------------------------------------------------------
\section{Conclusion}
%-------------------------------------------------------------------------------
\label{sec:conclusion}
\toolname is a new \pfour test oracle that automatically generates \inout tests for arbitrary \pfour targets. It uses whole-program semantics, taint-tracking, concolic execution, and path selection strategies to model the behavior of the \pfour program and generate tests that achieve coverage. \toolname is intended to be a resource for the entire \pfour community. It already supports \inout test generation for three open-source \pfour targets and several extensions for closed-source targets are in development. By designing it as a target-independent, extensible platform, we hope that \toolname will be well-positioned for long-term success. Moreover, since \toolname is a back end of \pfourc, it should be easy for developers to build on our tool, lowering the barrier of adoption.

As \toolname is an open-source tool, we welcome contributions from the broader community to improve and extend its functionality. For example, two common community requests are to extend \toolname with the ability (i) to generate arbitrarily many entries per table and (ii) produce tests with a state-preserving sequence of \inout packets. In the future, to further validate \toolname's generality, we would like to complete \toolname extensions for the P4-DPDK SoftNIC target and the open-source PSA~\cite{psa_spec} target for NIKSS~\cite{ubpf}, as well as proprietary SmartNICs~\cite{ipu,pensando_p4,nvidia_doca_sdk}. We also intend to develop additional \pfour validation tools based on \toolname's framework which apply ideas from software testing in the networking domain---\eg random program generation, mutation testing, and incremental testing. We are also interested in network-specific coverage notions---\eg for parsers, tables, actions, etc.

Software testing is always important, but testing the packet processing programs that power our network infrastructure, processing billions of packets per second, is especially important. In time, there will inevitably be better approaches than \toolname for generating high-quality tests for packet processing systems. The \toolname framework can serve as a vehicle for prototyping these approaches, and for integrating them into the  \pfour ecosystem. In the future, inspired by efforts from other communities~\cite{smtcomp,evaluating_fuzzing}, we envision having an open benchmark suite of standard test programs, control plane configurations, and various notions of coverage to standardize comparisons between different testing approaches---enabling more rapid progress for the whole community.

\Para{Ethics.} Our work on \toolname does not raise any ethical issues.
\section*{Acknowledgements}
\label{sec:ack}
We wish to thank Boris Beylin, Brad Burres, C\u{a}lin Ca\c{s}caval, Chris Dodd, Glen Gibb, Vladimir Gurevich, Changhoon Kim, Nick McKeown, Chris Sommers, Vladimir Still, and Edwin Verplanke for their support, detailed explanations of the semantics of different \pfour targets, and help with analyzing and classifying bugs. We are also grateful to Mihai Budiu, Andy Fingerhut, Dan Lenoski, Jonathan DiLorenzo, Ali Kheradmand, Steffen Smolka, and Hari Thantry for insightful conversations on  \pfour validation and the broader \pfour ecosystem. Finally, we would also like to thank Aurojit Panda, Tao Wang, our shepherd Sanjay Rao, and the participants of the Bellairs Workshop on Network Verification for valuable feedback on this work. This work was supported in part by NSF grant CNS-2008048.

%-------------------------------------------------------------------------------
\bibliographystyle{plain}
\bibliography{main}

\newpage
\onecolumn
\appendix
\section{Appendix}
Appendices are supporting material that has not been peer-reviewed.
\subsection{Target Implementation Details}
\label{app:target_quirks}

\begin{table*}[htb]
\centering
\begin{footnotesize}
%%%%%%%%%%%%%%%% TOFINO QUIRKS %%%%%%%%%%%%%%%%
\begin{tabularx}{\linewidth}{X}
\toprule
\textbf{\texttt{tna}/\texttt{t2a} target detail} \\
\midrule
\ding{43} \texttt{tna} has \textasciitilde48 extern functions and 6 programmable blocks~\cite{open_tofino}. \texttt{t2na} has over a 100 externs and 7 programmable blocks. \\
\ding{43} Tofino 2 adds a programmable block, the ghost thread. This block can update information related to queue depth in parallel to the packet traversing the program.  \\
\ding{43} In the Tofino parser, if a packet is too short to be read by an extern (extract/advance/lookahead) the packet is dropped, unless Tofino's ingress control reads the parser error variable. Then the packet header causing the exception is in an unspecified state~\cite[\S 5.2.1]{open_tofino}. \\
\ding{43} The packet that enters the Tofino parser is augmented with additional information, which needs to be modelled. Tofino 1 and 2 prepend metadata to the packet~\cite[\S 5.1]{open_tofino}. A 4-bytes Ethernet frame check sequence (FCS) is also appended. The parser can parse these values into \pfour data structures. \\
\ding{43} If the egress port variable is not set in the \pfour program, the packet is practically dropped (no unicast copy is made)~\cite[\S 5.1]{open_tofino}. \\
\ding{43} The value of the output port in Tofino matters. Some values are associated with the CPU port or recirculation, some are not valid, some forward to an output port. The semantics and validity of the ports can be configured~\cite[\S 5.7]{open_tofino}. \\
\ding{43} Tofino follows the Ethernet standard. Packets must have a minimum size of 64 bytes. Otherwise, the packet will be dropped~\cite[\S 7.2]{open_tofino}. The exception to this rule are packets injected from the Tofino CPU PCIe port.\\
\ding{43} The Tofino compiler provides annotations which can affect program semantics. Some annotations can alter the size of the \pfour metadata structure. If not handled correctly, this can affect the size of the output packet~\cite[\S 11]{open_tofino}. Another convenience annotation will initialize all otherwise random metadata to 0. \\
% \ding{43} Tofino has a notion of direct and indirect externs. Direct externs are attached to tables and can update register values in place. Indirect externs are only able to update register values from packet to packet. Reading and writing these externs on the same packet has no effect~\cite[\S 7.1]{open_tofino}. \\
\ding{43} The Tofino compiler removes all fields that are not read in the \pfour program from the egress metadata structure. This influences the size of the packet parsed by the egress parser. \\
\ding{43} Invalid access to header stacks in a parse loop will not cause a \texttt{StackOutOfBounds} error. Instead, execution transitions to the control with \texttt{PARSER\_ERROR\_CTR\_RANGE} set~\cite[\S 5.2.1]{open_tofino}. \\
\ding{43} Control plane keys in the Barefoot Runtime (Bfrt) may contain dollar signs (\$). When generating PTF/STF tests, these have to be replaced using a compiler pass. \\
\ding{43} Tofino has a metadata variable, which tells the traffic manager to skip egress processing entirely~\cite[\S 5.6]{open_tofino}. \\
\ding{43} Tofino 2 has a metadata variable, which instructs the deparser to truncate the emitted packet to the specified size. \\
\end{tabularx}
%%%%%%%%%%%%%%%% V1MODEL QUIRKS %%%%%%%%%%%%%%%%
\begin{tabularx}{\linewidth}{X}
\toprule
\textbf{\texttt{v1model} target detail} \\
\midrule
\ding{43} \texttt{v1model} has \textasciitilde26 extern functions and 6 programmable blocks~\cite{bmv2_ss}. \\
\ding{43} \bmv's default output port is 0~\cite{bmv2_ss}. \bmv drops packets when the egress port is 511. \\
\ding{43} When using Linux virtual Ethernet interfaces with \bmv, packets that are smaller than 14 bytes produce a curious sequence of hex output (02000000)~\cite{bmv2_outputs}. \\
\ding{43} \bmv supports a special technique to preserve metadata when recirculating a packet. Only the metadata that is annotated with field\_list and the correct index is preserved~\cite{bmv2_ss}. \\
\ding{43} \bmv supports the assume/assert externs which can cause \bmv to terminate abnormally~\cite{bmv2_assert_assume}. \\
\ding{43} \bmv's clone extern behaves differently depending on the location it was called in the pipeline. If recirculated in ingress, the cloned packet will have the values after leaving the parser and is directly sent to egress. If cloned in egress, the recirculated packet will have the values after it was emitted by the deparser~\cite{bmv2_ss}. \\
\ding{43} \bmv has an extern that takes the payload into account for checksum calculation. This means you always have to synthesize a payload for this extern~\cite{bmv2_ss}. \\
\ding{43} A parser error in \bmv does not drop the packet. The header that caused the error will be invalid and execution skips to ingress~\cite{bmv2_ss}. \\
\ding{43} All uninitialized variables are implicitly initialized to 0 or false in \bmv.  \\
\ding{43} Some \texttt{v1model} programs include \pfourconstraints, which limits the types of control plane entries that are allowed for a particular table.  \\
\ding{43} The table implementation in \bmv supports the priority annotation, which changes the order of evaluation of constant table entries.  \\
\end{tabularx}
%%%%%%%%%%%%%%%% EBPF QUIRKS %%%%%%%%%%%%%%%%
\begin{tabularx}{\linewidth}{X}
\toprule
\textbf{\texttt{ebpf\_model} target detail} \\
\midrule
\ding{43} \texttt{ebpf\_model} has 2 extern functions and 2 programmable blocks. \\
\ding{43} The eBPF target does not have a deparser that uses emit calls. It can only filter. \\
\ding{43} \texttt{extract} or \texttt{advance} have no effect on the size of the outgoing packet.  \\
\ding{43} A failing \texttt{extract} or \texttt{advance} in the eBPF kernel automatically drops the packet.  \\
\bottomrule
\end{tabularx}

\end{footnotesize}
\caption{A nonexhaustive collection of target implementation details that require \toolname's use of whole-program semantics to provide an accurate model. Where possible, we cited a source. Some details are not explicitly documented.}
\label{tbl:target_quirks}
\end{table*}

\subsection{Program Measurements}
\label{app:measurements}

\begin{table*}[!htb]
\centering
\begin{footnotesize}
\begin{tabular}{p{11.5em}|lll|lll|lll|lll} 
\toprule
Program  & \multicolumn{3}{l|}{\middleblock} & \multicolumn{3}{l|}{\upfour}  & \multicolumn{3}{l}{\tnasimpleswitch} & \multicolumn{3}{l}{\dash} \\ 
\midrule
\diagbox[innerwidth=11em]{Strategy}{Metric (Median)}   & \begin{tabular}[c]{@{}l@{}}Tests\end{tabular} & \begin{tabular}[c]{@{}l@{}}Time \\per test\end{tabular} & \begin{tabular}[c]{@{}l@{}}Total time\end{tabular} & \begin{tabular}[c]{@{}l@{}}Tests\end{tabular} & \begin{tabular}[c]{@{}l@{}}Time \\per test\end{tabular} & \begin{tabular}[c]{@{}l@{}}Total time\end{tabular} & \begin{tabular}[c]{@{}l@{}}Tests\end{tabular} & \begin{tabular}[c]{@{}l@{}}Time \\per test\end{tabular} & \begin{tabular}[c]{@{}l@{}}Total time\end{tabular} & \begin{tabular}[c]{@{}l@{}}Tests\end{tabular} & \begin{tabular}[c]{@{}l@{}}Time \\per test\end{tabular} & \begin{tabular}[c]{@{}l@{}}Total time\end{tabular} \\ 
\midrule
% MIDDLEBLOCK                                                 % UP4                                 % TNA_SIMPLE_SWITCH                     % DASH_PIPELINE
\dfscov      & 25105 & \textasciitilde{}0.05 & 1321.4s  & 12932 & \textasciitilde{}.06s & 726.34s   & * & \textasciitilde{}0.07 & * & *     & \textasciitilde{}0.06 & * \\
\randcov     & 956 & \textasciitilde{}.08s & 80.92s        & 2463 & \textasciitilde{}.07s & 169.3s  & * & \textasciitilde{}0.09 & * & *     & \textasciitilde{}0.13 & * \\
\maxcov      & 86 & \textasciitilde{}.17s & 11.41s          & 3581 & \textasciitilde{}.06s & 242.6s & 4612 & \textasciitilde{}0.12 & 555.24 & 63 & \textasciitilde{}0.41 & 21.86s \\
\bottomrule
\end{tabular}
\end{footnotesize}
\caption{Path selection results for 100\% statement coverage on representative \pfour programs for 10 different seeds. "*" indicates that the strategy did not achieve 100\% coverage within 60 minutes.}
\label{tbl:strategies}
\end{table*}

%%%%%%%%%%%%%%%%%%%%%%%%%%%%%%%%%%%%%%%%%%%%%%%%%%%%%%%%% Middleblock Tests
\begin{table*}[!htb]
\begin{footnotesize}
\centering
\begin{tabular}{l|cccccc} 
\toprule
\begin{tabular}[c]{@{}l@{}}Applied\\precondition\end{tabular} & None   & \begin{tabular}[c]{@{}l@{}}Fixed-Size\\Packet\end{tabular} & \pfourconstraints & \begin{tabular}[c]{@{}l@{}}\pfourconstraints\\Fixed-Size Packet\end{tabular} & \begin{tabular}[c]{@{}l@{}}\pfourconstraints, Fixed-\\Size IPv4 Packet\end{tabular} & \begin{tabular}[c]{@{}l@{}}\pfourconstraints, Fixed-Size\\IPv4-TCP Packet\end{tabular}  \\ 
\midrule
Valid test paths    & 146784 & 83784 & 74472 & 42486 & 28216 & 7054 \\
\textbf{Reduction}           & \textbf{0\%}    & \textbf{\textasciitilde43\%} & \textbf{\textasciitilde49\%} & \textbf{\textasciitilde71\%} & \textbf{\textasciitilde81\%} & \textbf{\textasciitilde95\%} \\
\bottomrule
\end{tabular}
\caption{Effect of preconditions on the number of tests generated for \middleblock. Fixed packet size is 1500B.}
\label{tbl:middleblock_reduction}
\end{footnotesize}
\end{table*}

% %%%%%%%%%%%%%%%% Bug details %%%%%%%%%%%%%%%%
\begin{table*}[!htb]
\centering
\begin{footnotesize}
\begin{tabularx}{\linewidth}{lllX}
\toprule
Bug label & Type & Bug description\\
\midrule
\href{https://github.com/p4lang/PI/issues/585}{\path{p4lang/PI/issues/585}} & Exception & The open-source P4Runtime server has incomplete support for the \texttt{p4runtime\_translation} annotation.\\
\href{https://github.com/p4lang/behavioral-model/issues/1179}{\path{p4lang/behavioral-model/issues/1179}} & Exception & \bmv crashes when trying to add entries for a shared action selector.\\
\href{https://github.com/p4lang/p4c/issues/3423}{\path{p4lang/p4c/issues/3423}} & Exception & \bmv crashes when accessing a header stack with an index that is out of bounds.\\
\href{https://github.com/p4lang/p4c/issues/3514}{\path{p4lang/p4c/issues/3514}} & Exception & The STF test back end is unable to process keys with expressions in their name.\\
% \href{https://github.com/p4lang/p4c/issues/3509}{\path{p4lang/p4c/issues/3509}} & Exception & The compiler did not correctly transform a varbit \texttt{extract} call with an expression as second argument.\\
\href{https://github.com/p4lang/p4c/issues/3429}{\path{p4lang/p4c/issues/3429}} & Exception & The output by the compiler was using an incorrect operation to dereference a header stack.\\
\href{https://github.com/p4lang/p4c/issues/3435}{\path{p4lang/p4c/issues/3435}} & Exception & Actions, which are missing their ``name'' annotation, cause the STF test back end to crash.\\
% \href{https://github.com/p4lang/p4c/issues/3508}{\path{p4lang/p4c/issues/3508}} & Exception & A second instance where the compiler was using the wrong operation to manipulate header stacks.\\
% \href{https://github.com/p4lang/p4c/issues/3862}{\path{p4lang/p4c/issues/3862}} & Exception & The compiler should have flattened a header union input for \texttt{emit} calls.\\
\href{https://github.com/p4lang/p4c/issues/3620}{\path{p4lang/p4c/issues/3620}} & Exception & \bmv can not process structure members with the same name.\\
\href{https://github.com/p4lang/p4c/issues/3490}{\path{p4lang/p4c/issues/3490}} & Wrong code & The compiler swallowed the \texttt{table.apply()} of a switch case, which led to incorrect output.\\
\bottomrule
\end{tabularx}
\caption{\bmv bugs found by \toolname. }
\label{tbl:bug_list}
\end{footnotesize}
\end{table*}

\subsection{Packet-Sizing}
\label{app:packet_sizing}
%---- FIGURE ----%
\begin{figure*}[!htb]
% FIGURE 1
\begin{subfigure}[b]{0.49\columnwidth}
\begin{lstlisting}[style=P4Style]
parser IngressParser(...) {
    state start {
        pkt.extract(ingress_meta);
        pkt.extract(hdr.eth);
        pkt.extract(hdr.ipv4);
    }
}
control IngressControl(...) {
    apply {}
}
control IngressDeparser(...) {
    apply {
        pkt.extract(ingress_meta);
        pkt.extract(hdr.eth);
        pkt.extract(hdr.ipv4);
    }
}
parser EgressParser(...) {
    state start {
        pkt.extract(egress_meta);
        pkt.extract(hdr.eth);
    }
}
control EgressControl(...)
    apply {}
}
control EgressDeparser(...) {
    apply {
        pkt.extract(hdr.eth);
    }
}
Pipeline(
    IngressParser(), Ingress(), IngressDeparser(),
    EgressParser(), Egress(), EgressDeparser()
    ) pipe;
Switch(pipe) main;

\end{lstlisting}
    \caption{Extern sequence manipulating Ethernet and IPv4 headers.}
    \label{fig:packet_sizing_a}
\end{subfigure}
\hfill
% FIGURE 2
\begin{subfigure}[b]{0.5\columnwidth}
    \centering
    \includegraphics[width=\columnwidth]{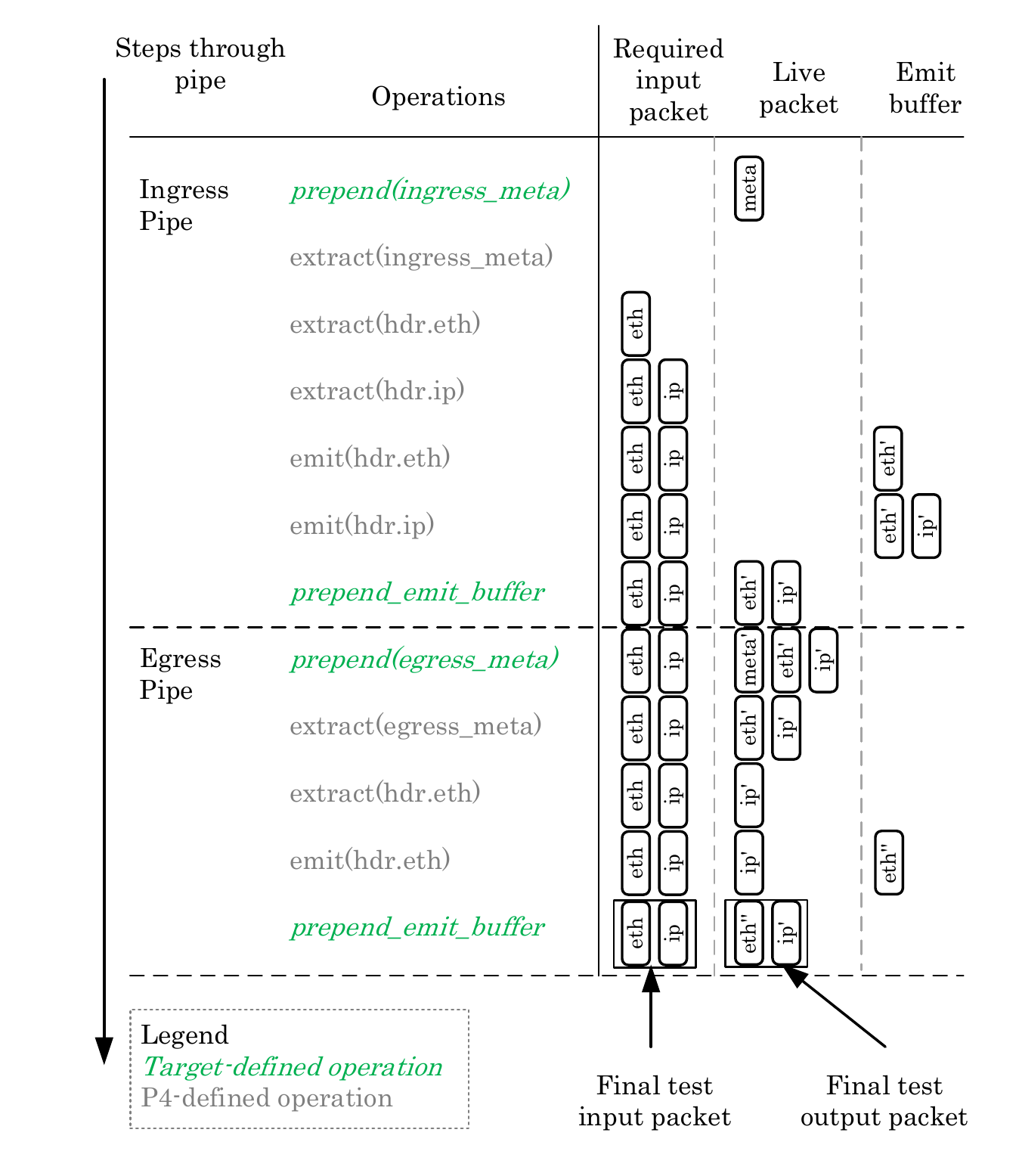}
    \caption{Change in the packet sizing variables as \toolname steps through the program. Each block corresponds to a \pfour header. }
    \label{fig:packet_sizing_b}
\end{subfigure}
% \vspace{-2em}
\caption{Packet-sizing for a Tofino program. }
\label{fig:packet_sizing}
\end{figure*}
%---- FIGURE ----%

%%%%%%%%%%%%%%%%%%%%%%%%%%%%%%%%%%%%%%%%%%%%%%%%%%%%%%%%%%%%%%%%%%%%%%%%%%%%%%%%
\end{document}